\documentclass{amsart}

\usepackage{cmap}
\usepackage[T1]{fontenc}

\usepackage{amsmath,amsfonts,amssymb,amsthm}
\usepackage{graphicx}
\usepackage{thmtools, thm-restate}

\usepackage{hyperref}

\usepackage[utf8]{inputenc}
\usepackage[english]{babel}

\declaretheorem[name=Theorem,style=plain,numberwithin=section]{thm}
\declaretheorem[name=Lemma,style=plain,sibling=thm]{lem}
\declaretheorem[name=Corollary,style=plain,sibling=thm]{cor}
\declaretheorem[name=Proposition,style=plain,sibling=thm]{prop}

\declaretheorem[name=Definition,style=definition,sibling=thm]{defn}

\declaretheorem[name=Remark,style=definition,sibling=thm]{remark}

\newcounter{proofpart}
\renewcommand\theproofpart{\Roman{proofpart}}
\newenvironment{proofpart}[1]{\stepcounter{proofpart}\vspace{0.5pc}\noindent{\textbf{Step \theproofpart: #1\\}}}{}
\makeatletter
\@addtoreset{proofpart}{thm}
\@addtoreset{proofpart}{lem}
\@addtoreset{proofpart}{prop}
\@addtoreset{proofpart}{cor}
\makeatother

\newenvironment{claim}[1]{\refstepcounter{proofpart}\vspace{0.5pc}\noindent{\textbf{Step \theproofpart: #1\\}}}{}

\usepackage[mathscr]{euscript}

\hypersetup{
    colorlinks=false,
    pdfborder={0 0 0},
}

\newcommand{\R}{\mathbb R}
\newcommand{\Z}{\mathbb Z}
\newcommand{\N}{\mathbb N}

\newcommand{\im}{\operatorname{im}}
\newcommand{\tr}{\operatorname{tr}}
\newcommand{\id}{\operatorname{id}}

\newcommand{\edge}{\operatorname{edge}}

\newcommand{\deriv}[2]{\frac{d#1}{d#2}}
\newcommand{\partderiv}[2]{\frac{\partial #1}{\partial #2}}
\newcommand{\partdderiv}[2]{\frac{\partial^2 #1}{\partial #2^2}}

\newcommand{\Ric}{\operatorname{Ric}}

\newcommand{\Al}{_{\mathcal Al}}

\newcommand{\quotient}[2]{\left. #1 \right/ #2}

\title{Smoothness of compact horizons}
\author{Eric Larsson}
\address{%
Department of Mathematics,\\
KTH Royal Institute of Technology\\
SE-100 44 Stockholm\\
Sweden
}
\email{ericlar@kth.se}

\begin{document}

\maketitle

\begin{abstract}
We prove that compact Cauchy horizons in a smooth spacetime satisfying the null energy condition are smooth.
As an application, we consider the problem of determining when a cobordism admits Lorentzian metrics with certain properties. In particular, we prove a result originally due to Tipler without the smoothness hypothesis necessary in the original proof.
\end{abstract}

\section*{Introduction}
An intriguing question in the theory of general relativity is that of \emph{topology change}: Is it possible for a spacelike slice of spacetime at one time to have a different topology than that of a spacelike slice at some other time? One way of making this question precise is through the concept of a \emph{Lorentzian cobordism} (see \autoref{Definition of Lorentzian cobordism}); that is, a spacetime whose boundary consists of disjoint spacelike submanifolds. The question whether topology change is possible can then be interpreted as the question of whether physically interesting nontrivial Lorentzian cobordisms exist. For this question to be interesting the cobordism needs to have some compactness property. We will consider both the case when the cobordism is compact, and the case when the cobordism has the weaker property of causal compactness (see \autoref{Definition of causal compactness}). 

The existence of Lorentzian cobordisms when no geometrical conditions are imposed is essentially a problem of differential topology. It is equivalent to the existence of a cobordism with a vector field with a prescribed direction at the boundary, and the problem of characterizing pairs of manifolds which are cobordant in this sense was considered by Reinhart \cite{Reinhart63}.

When geometrical conditions are imposed, it is significantly more difficult to construct Lorentzian cobordisms which are not diffeomorphic to $S \times [0, 1]$ for some manifold $S$.
There are two classical results about the non-existence of nontrivial Lorentzian cobordisms under certain hypotheses: In 1967 it was shown by Geroch \cite{Geroch67} that nontrivial Lorentzian cobordisms can exist only if they contain closed timelike curves, and in 1977 it was shown by Tipler \cite{Tipler77} that nontrivial Lorentzian cobordisms satisfying certain energy conditions cannot exist. 

In proving Tipler's theorem, one works with a compact Cauchy horizon, and the question arises which regularity a Cauchy horizon has. It was shown by Chru\'sciel and Galloway in \cite{ChruscielGalloway98} that Cauchy horizons can be far from smooth. In \cite[Section~IV]{BeemKrolak98} Beem and Kr\'olak proved that a compact ''almost everywhere $C^2$'' horizon satisfying the null energy condition is everywhere $C^1$. In \cite[Section~4]{BudzynskiKondrackiKrolak01} by Budzy\'nski, Kondracki, and Kr\'olak it was asked whether \emph{compact} Cauchy horizons are always smooth. This question was answered in the negative in \cite{BudzynskiKondrackiKrolak03} by the same authors, where it was also suggested that an energy condition might be sufficient to conclude that a compact Cauchy horizon is smooth.

The main result of this paper is \autoref{Compact horizons are smooth}, where we prove that compact Cauchy horizons in a spacetime satisfying the null energy condition are smooth, thereby significantly generalizing the theorem in \cite[Section~IV]{BeemKrolak98}, and providing an answer to the the question raised in \cite[Section~4]{BudzynskiKondrackiKrolak03} of whether compact horizons which satisfy energy conditions are smooth. We then apply this theorem to obtain a complete proof of Tipler's theorem. It has been known for some time (see for instance \cite[Section~1]{ChruscielGalloway98}) that the proof of Tipler's result in \cite{Tipler77} makes an implicit smoothness assumption. The proof uses arguments from the proof of the Hawking Singularity Theorem~\cite[p.\,295-298]{HawkingEllis}, and the same implicit assumption can be found there as well. 
A similar oversight was made in the original proof of the Hawking Area Theorem, and has since been corrected by Chru\'sciel, Delay, Galloway and Howard \cite{CDGH01}. Significant work was necessary to fill in the gaps in the proof of the Hawking Area Theorem, and the proof in \cite{CDGH01} is technical. Fortunately, their methods may be adapted to the setting of Tipler's theorem and we do so in this paper in order to present a more careful proof of the nonexistence of Lorentzian cobordisms which satisfy certain energy conditions. 

\autoref{Compact horizons are smooth} is also interesting in relation to the papers \cite{IsenbergMoncrief85}, \cite{MoncriefIsenberg83}, and \cite{MoncriefIsenberg08} by Isenberg and Moncrief. In the first two, it is shown that analytic compact null hypersurfaces with certain properties admit a null Killing vector field. \autoref{Compact horizons are smooth} allows us to drop the hypothesis that the hypersurface is analytic, and replace it with the hypothesis that the null energy condition holds and that the hypersurface is a Cauchy horizon. In fact, as is discussed in \autoref{Stronger results}, it is not necessary that the hypersurface is a Cauchy horizon, provided its generators are complete to the past.

Appendix \ref{Geometric measure theory} contains a summary of results from geometric measure theory.

\subsection*{A comment}
After these results were published as part of the author's Master's thesis \cite{LarssonThesis}, similar work by Minguzzi (\cite{MinguzziArea} and \cite{MinguzziCompleteness}) has appeared.
Following concerns raised in \cite{MinguzziCompleteness} about the existence of a timelike vector field $V$ satisfying $\nabla_V V = 0$ in a neighborhood of a Cauchy horizon, a construction of a vector field having this property on a sufficiently large set has been added to \autoref{H+complete}.

\section{Smoothness of compact Cauchy horizons}\label{Smoothness of compact Cauchy horizons}
The purpose of this section is to prove \autoref{Compact horizons are smooth}. We begin by stating and proving some properties of $C^2$ null hypersurfaces in \autoref{C2 null hypersurfaces}. We then define the concept of a ''horizon'' and summarize some previously known results about horizons in \autoref{Structure of horizons}. Finally, we prove the smoothness theorem (\autoref{Compact horizons are smooth}) in \autoref{A smoothness theorem}.
\subsection{\texorpdfstring{$C^2$}{C\^{}2} null hypersurfaces}\label{C2 null hypersurfaces}
\subsubsection{The null Weingarten map}\label{The null Weingarten map}
In the following section we summarize properties of $C^2$ null hypersurfaces which we will need later. For details, see \cite{GallowayNullGeometry}, \cite{Kupeli87}, \cite[Section~II.1]{Galloway00} or \cite[Appendix A]{CDGH01}.

A null hypersurface $\mathscr H$ in a spacetime $M$ is characterized by the fact that each tangent space $T_p\mathscr H$ contains a unique (up to scaling) null vector $K_p$. The tangent space $T_p\mathscr H$ then consists of those vectors of $T_pM$ which are orthogonal to $K_p$. This means that any normal vector field $K$ of $\mathscr H$ consists entirely of null vectors. We will call the integral curves of these vector fields \emph{generators} of $\mathscr H$. By \cite[Proposition 3.1]{GallowayNullGeometry} these generators (when given a suitable parametrization) are geodesics.
By straightforward computations it holds that
\[\langle X, Y \rangle = \langle X', Y' \rangle \quad \text{and} \quad \langle \nabla_X K, Y \rangle = \langle \nabla_{X'} K, Y' \rangle\]
whenever $X, Y \in T_p\mathscr H$ and $X - X' = \lambda_1 K$ and $Y - Y' = \lambda_2 K$ for some real numbers $\lambda_1, \lambda_2$. Inspired by this, we work instead with the quotient space $\quotient{T_p\mathscr H}{\R K}$. 
This quotient is independent of the particular choice of null vector field $K$, since all such vector fields differ only by scaling. We now define the \emph{null Weingarten map} of $\mathscr H$ with respect to $K$ by
\[b_K \colon \quotient{T\mathscr H}{\R K} \to \quotient{T\mathscr H}{\R K},\]
\[b_K(\overline X) = \overline{\nabla_X K}.\]
This map is not independent of the particular choice of null vector field $K$. However, if $f$ is a smooth function without zeros then $b_{fK} = fb_K$ since $K$ is null. Note that if $\mathscr H$ is $C^2$, then $K$ can be chosen $C^1$ so that $b_K$ is continuous.
Since all our spacetimes are time-oriented we may restrict attention to future-directed null vector fields $K$. This means that we can associate to each null hypersurface a family of null Weingarten maps which differ only by positive scaling. 
Since $K$ is null the spacetime metric induces an inner product on $\quotient{T\mathscr H}{\R K}$. Using this inner product, we may define the \emph{null second fundamental form} of $\mathscr H$ with respect to $K$ by
\[B_K(\overline X, \overline Y) = \langle b_K(\overline X), \overline{Y} \rangle.\]
A straightforward computation shows that $B_K$ is symmetric. 
We will need the following theorem, a proof of which can be found in \cite[Theorem~30]{Kupeli87}.
\begin{thm}\label{Vanishing null second fundamental form implies totally geodesic hypersurface}
Let $\mathscr H$ be a smooth null hypersurface in a spacetime $M$. Then the null second fundamental form of $\mathscr H$ is identically zero if and only if $\mathscr H$ is a totally geodesic submanifold of $M$.
\begin{remark}
The theorem as stated in \cite[Theorem~30]{Kupeli87} applies to null submanifolds in general, regardless of codimension, and so requires the submanifold to be ''irrotational''. This condition is automatically satisfied for null hypersurfaces.
\end{remark}
\end{thm}
Finally, we define the \emph{null mean curvature} $\theta_K$ of a null hypersurface with respect to a null vector field $K$ as the trace of the null Weingarten map:
\[\theta_K = \tr b_K.\]
Recall that if $K' = \lambda K$ is another null vector field then $b_{K'} = \lambda b_K$. Hence $\theta_{K'} = \lambda \theta_K$. This means that the sign of $\theta_K$ is independent of the particular future-directed null vector field $K$ used to compute $\theta_K$. We will sometimes omit the vector field $K$ from the notation.

Recall that the integral curves of a null vector field $K$ are reparametrizations of geodesics. If $K$ is chosen to agree with $\dot \gamma$ of a geodesic segment $\gamma$ with affine parameter $s$, and $b(s)$ is the family of null Weingarten maps with respect to $K$ along $\gamma$, then
\begin{equation}\label{Riccati equation for null Weingarten map}
\dot b + b^2 + \widetilde R = 0.
\end{equation}
Here $\dot b$ denotes the derivative of $b$ along $\gamma$, and $\widetilde R$ denotes the fiberwise endomorphism $\widetilde R \colon \quotient{T\mathscr H}{\R K} \to \quotient{T\mathscr H}{\R K}$ defined from the Riemann curvature tensor $R$ by letting ${\widetilde R(\overline X) = \overline{R(X, \dot\gamma)\dot\gamma}}$. Note that it is not obvious that the derivative $\dot b$ exists, since $b$ is a priori only continuous. A proof of the fact that the derivative does exist and satisfies equation \eqref{Riccati equation for null Weingarten map} can be found in \cite[Proposition A.1]{CDGH01}.

From equation \eqref{Riccati equation for null Weingarten map} one can derive the \emph{Raychaudhuri equation}. In particular, one may derive a certain differential inequality for the null mean curvature. Let $b$ be the null Weingarten map of a $C^2$ null hypersurface with respect to a future-directed null vector field $K$ (scaled to give an affine parametrization of an integral curve), and let $\theta$ denote the trace of $b$. Let $S = b - \frac{\theta}{n-2}\id$. Then the trace of $b^2$ is $\tr b^2 = \theta^2/(n-2) + \tr(S^2)$ so taking the trace of equation \eqref{Riccati equation for null Weingarten map} yields
\begin{equation}\label{Raychaudhuri equation}
\dot \theta + \frac{\theta^2}{n-2} + \tr(S^2) + \Ric(K, K) = 0.
\end{equation}
Since $b$ and $\id$ are self-adjoint, so is $S$. Hence $\tr(S^2) \geq 0$ so
\begin{equation}\label{Raychaudhuri inequality}
\dot \theta + \frac{\theta^2}{n-2} + \Ric(K, K) \leq 0.
\end{equation}
This is the differential inequality we will use later.

\subsubsection{Generator flow on \texorpdfstring{$C^2$}{C\^{}2} null hypersurfaces}\label{Generator flow on C2 null hypersurfaces}
A null vector field on a $C^2$ null hypersurface gives rise to a family of local diffeomorphisms which move points along the vector field. The integral curves of such a vector field are called generators, and we will refer to such a flow as a \emph{generator flow}. The generator flow for time $t$ will be denoted $\beta_t$. Given a Riemannian metric $\sigma$ on $M$ with volume form $\omega$ on the null hypersurface, the \emph{Jacobian determinant} $J(\beta_t)$ with respect to $\sigma$ is the real valued function defined by $(\beta_t)^*\omega = J(\beta_t)\omega$.
In this section, we will show that the Jacobian determinant of a generator flow with respect to Riemannian metrics of a certain form is related to the null mean curvature of the hypersurface. We choose to work with a past-directed vector field $T$ since this is the case in which we will apply the lemma.
\begin{lem}\label{Lie derivative computation}
Let $\mathscr H$ be a $C^2$ null hypersurface in a spacetime $(M, g)$ of dimension $n+1$. Let $V$ be an arbitrary unit timelike vector field on $M$, and define a Riemannian metric $\sigma$ on $M$ by
\[\sigma(X, Y) = g(X, Y) + 2g(X, V)g(Y, V).\]
Let $T$ denote the unique past-directed lightlike $\sigma$-unit vector field on $\mathscr H$ and let $\omega$ denote the $\sigma$-volume form induced on $\mathscr H$. Let $\theta$ denote the null mean curvature of $\mathscr H$ with respect to the future-directed null vector field $-T$. Then the Lie derivative of $\omega$ with respect to $T$ is $\mathcal L_T \omega = -\theta\omega$.
\begin{proof}
Choose some point $p \in \mathscr H$ at which to evaluate $\mathcal L_T \omega$. Let $e_1, e_2, \ldots, e_n$ be a $g$-orthogonal basis for $T_p\mathscr H$ such that
\begin{itemize}
\item $e_1 = T_p$,
\item $g(e_i, V) = 0$ for $i = 2, 3, \ldots, n$,
\item $g(e_i, e_i) = 1$ for $i = 2, 3, \ldots, n$.
\end{itemize}
Recall that integral curves of null vector fields on null hypersurfaces are geodesic segments. Let $\gamma$ be a segment of the integral curve of $T$ through $p$ with an affine parametrization. 
Extend the basis $e_1, \ldots, e_n$ along $\gamma$ by letting
\begin{itemize}
\item $e_1 = T$,
\item $\nabla_{e_1} e_i = 0$ for $i = 2, 3, \ldots, n$.
\end{itemize}
Here $\nabla$ denotes covariant derivative with respect to $g$. Note that we do not yet know that $(e_i)_{i = 1}^n$ is a frame for $\mathscr H$, since we first need to show that the $e_i$ are tangent to $\mathscr H$. This will follow from the properties below.
Direct computations show that $(e_i)_{i = 1}^n$ have the following properties on the image of $\gamma$.
\begin{itemize}
\item $g(e_1, V) = 1/\sqrt{2}$.
\item $g(e_1, e_i) = 0$.
\item $g(e_i, e_j) = \delta_{ij}$ for $i,j = 2,3,\ldots,n$.
\item $\sigma(e_1, e_i) = \sqrt{2}g(e_i, V)$ for $i = 2, 3, \ldots, n$.
\item $\sigma(e_i, e_j) = \delta_{ij} + 2g(e_i, V)g(e_j, V)$ for $i,j = 2, 3, \ldots, n$.
\end{itemize}

\begin{claim}{$\det(\sigma(e_i, e_j)) = 1$}
Let $a_i = \sigma(e_i, V)$ for $i = 1, 2, \ldots, n$. By the previous claims the matrix $A$ with entries $\sigma(e_i, e_j)$ can then be written as
\[A = \begin{pmatrix}
1 & \sqrt{2} a_2 & \sqrt{2} a_3 & \cdots\\
\sqrt{2}a_2 & 1 + 2a_2^2 & 2 a_2 a_3 & \cdots\\
\sqrt{2}a_3 & 2a_2a_3 & 1 + 2 a_3^2 & \cdots\\
\vdots & \vdots & \vdots & \ddots
\end{pmatrix}.\]
Let
\[B = \begin{pmatrix}
1 & 0 & 0 & 0 & \cdots\\
-\sqrt{2}a_2 & 1 & 0 & 0 & \cdots\\
-\sqrt{2}a_3 & 0 & 1 & 0 & \cdots\\
-\sqrt{2}a_4 & 0 & 0 & 1 & \cdots\\
\vdots & \vdots & \vdots & \vdots & \ddots
\end{pmatrix}.\]
Then $\det B = 1$. Moreover
\[BA = \begin{pmatrix}
1 & \sqrt{2} a_2 & \sqrt{2} a_3 & \cdots\\
0 & 1 & 0 & \cdots\\
0 & 0 & 1 & \cdots\\
\vdots & \vdots & \vdots & \ddots
\end{pmatrix}.\]
Hence $\det(BA) = 1$. This means that
\[\det A = \frac{\det(BA)}{\det B} = 1,\]
proving the claim.
\end{claim}

\begin{proofpart}{Computation of $\mathcal L_T \omega$}
The volume form $\omega$ induced on $\mathscr H$ by the Riemannian metric $\sigma$ can be expressed in the frame $e_1, e_2, \ldots, e_n$ as
\[\omega = \sqrt{\det(\sigma(e_i, e_j))} e^1 \wedge e^2 \wedge \cdots \wedge e^n\]
where the $e^i$ are the covectors defined by $e^i(e_i) = 1$ and $e^i(e_j) = 0$ for $i \neq j$. By the previous claim the determinant is equal to $1$, so
\[\omega = e^1 \wedge e^2 \wedge \cdots \wedge e^n\]
on all of $\gamma$.
We will use Cartan's formula to compute the Lie derivative $\mathcal L_T \omega$, so we need to extend the frame $e_i$ to a neighborhood of $\gamma$. Extend $e_1, \ldots, e_n$ to a frame such that $e_1 = T$. Extend the dual frame $e^1, \ldots, e^n$ in the natural way by letting $e^i(e_i) = 1$ and $e^i(e_j) = 0$ for $i \neq j$. Rescale $e_n$ if necessary so that $\omega = e^1 \wedge e^2 \wedge \cdots \wedge e^n$ holds everywhere. 
We will now use this frame to compute $\mathcal L_T \omega$ at the point $p$.
By Cartan's formula
\[\mathcal L_T \omega = i_T d\omega + d(i_T \omega).\]
Since $\omega$ is an $n$-form on an $n$-manifold we have $d\omega = 0$. Hence
\[\mathcal L_T \omega = d(i_T \omega).\]
Since $\omega = e^1 \wedge e^2 \wedge \cdots \wedge e^n$
\[d(i_T \omega) = d(e^1(T) e^2 \wedge \cdots \wedge e^n) = d(e^1(e_1) e^2 \wedge \cdots \wedge e^n) = d(e^2 \wedge \cdots \wedge e^n).\]
Hence
\[\mathcal L_T \omega = \sum_{k = 2}^n (-1)^k e^2 \wedge \cdots \wedge e^{k-1} \wedge de^k \wedge e^{k+1} \wedge \cdots \wedge e^n.\]
We now compute $de^k$, or rather the part of $de^k$ which does not contain any $e^j$ for $j \notin \{1, k\}$; all such terms are annihilated when we insert this expression into the large wedge product above. Since $de^k$ is a two-form this means that only one of its terms, $(de^k)(e_1, e_k) e^1 \wedge e^k$, is interesting. Now
\[(de^k)(e_1, e_k) = e_1(e^k(e_k)) - e_k(e^k(e_1)) - e^k([e_1, e_k]) = - e^k([e_1, e_k])\]
since $e^k(e_k) = 1$ and $e^k(e_1) = 0$ close to $\gamma$. 
We express the Lie bracket, evaluated at the point $p$, using the spacetime metric $g$ as
\[(de^k)(e_1, e_k) = - e^k([e_1, e_k]) = -e^k(\nabla_{e_1} e_k - \nabla_{e_k} e_1) = e^k(\nabla_{e_k} e_1).\]
Recall that $\nabla$ denotes covariant derivative with respect to $g$.
We have used that $e_2, \ldots, e_n$ have been chosen such that $\nabla_{e_1} e_k = 0$ for all $k$. We now know that
\[de^k = (e^k(\nabla_{e_k} e_1))e^1 \wedge e^k + \ldots\]
where the dots signify terms containing some $e^j$ for $j \notin \{1, k\}$. At the point $p$ it then holds that
\[\begin{split}
(-1)^k & e^2 \wedge \cdots \wedge e^{k-1} \wedge de^k \wedge e^{k+1} \wedge \cdots \wedge e^n\\
&= (-1)^k e^2 \wedge \cdots \wedge e^{k-1} \wedge (e^k(\nabla_{e_k} e_1))e^1 \wedge e^k \wedge e^{k+1} \wedge \cdots \wedge e^n\\
&= (-1)^k(-1)^{k-2} (e^k(\nabla_{e_k} e_1)) e^1 \wedge e^2 \wedge \cdots \wedge e^{k-1} \wedge e^k \wedge e^{k+1} \wedge \cdots \wedge e^n\\
&= (e^k(\nabla_{e_k} e_1)) e^1 \wedge e^2 \wedge \cdots \wedge e^n
\end{split}\]
where the additional factor of $(-1)^{k-2}$ is due to commuting $e^1$ with the $e^2, \ldots, e^{k-1}$. Hence
\[\mathcal L_T \omega = \sum_{k = 2}^n (e^k(\nabla_{e_k} e_1)) e^1 \wedge e^2 \wedge \cdots \wedge e^n = \sum_{k = 2}^n (e^k(\nabla_{e_k} e_1)) \omega.\]
Since $e_1 = T$ and $e_2, \ldots, e_n$ are $g$-orthonormal and $g$-orthogonal to $e_1$,
\[\sum_{k = 2}^n e^k(\nabla_{e_k} e_1) = \sum_{k = 2}^n g(e_k, \nabla_{e_k} e_1) = -\sum_{k = 2}^n g(e_k, \nabla_{e_k} (-T)).\]
Recall from \autoref{The null Weingarten map} that the quotient $\quotient{T_p\mathscr H}{\R T}$ has an inner product induced by $g$ such that the image of $(e_i)_{i= 2}^n$ under the projection $T_p\mathscr H \to \quotient{T_p\mathscr H}{\R T}$ forms an orthonormal basis. This means that $\sum_{k = 2}^n g(e_k, \nabla_{e_k} (-T))$ is the trace of the null Weingarten map $b_{-T}$ defined in \autoref{The null Weingarten map}. This trace is the null mean curvature $\theta$ of $\mathscr H$ with respect to $-T$. 
We can then conclude that
\[\mathcal L_T \omega = -\theta \omega\]
at $p$. Since $p$ was arbitrary, this completes the proof.
\end{proofpart}
\end{proof}
\end{lem}
The proof of the following lemma essentially consists of integrating the Lie derivative $\mathcal L_T\omega$ to relate the null mean curvature $\theta$ to the Jacobian determinant of the generator flow. 
\begin{lem}\label{Integral expression for J}
Let $\mathscr H$ be a $C^2$ null hypersurface in a spacetime $(M, g)$. Let $\sigma$ be a Riemannian metric on $M$ of the form
\[\sigma(X, Y) = g(X, Y) + 2g(X, V)g(Y, V)\]
for some $g$-unit timelike vector field $V$ on $M$. Let $T$ be the unique past-directed $\sigma$-unit null vector field on $\mathscr H$, and let $\theta$ be the null mean curvature of $\mathscr H$ with respect to $-T$. 
Fix $t > 0$ and let $\beta_t \colon \mathscr H \to \mathscr H$ denote the flow along $T$ for time $t$ (whenever defined). Suppose that $p$ is such that $\beta_s(p)$ is defined for all $s \in [0, t]$. Let $J(\beta_t)$ denote the Jacobian determinant of $\beta_t$ with respect to $\sigma$. Then
\[J(\beta_t)(p) = \exp\left(-\int_0^t \theta(\beta_s(p))\,ds\right).\]
\begin{proof}
Let $\omega$ denote the volume form of $\sigma$. The Jacobian determinant $J(\beta_t)(p)$ is characterized by
\[\beta_t^*(\omega_{\beta_t(p)}) = J(\beta_t)(p) \omega_p.\]
For simpler notation, let $\alpha \colon [0, t] \to \R$ denote the function $\alpha(s) = J(\beta_s)(p)$.
Note that $\alpha(0) = 1$ since $\beta_0$ is the identity. 
By \cite[Proposition 12.36]{Lee} it holds that
\[\left.\deriv{}{\tau}\right|_{\tau = s} \beta_{\tau}^*(\omega_{\beta_\tau(p)}) = \beta_s^*(\mathcal L_T \omega_{\beta_s(p)}).\]
Since $\beta_{\tau}^*(\omega_{\beta_\tau(p)}) = \alpha(\tau)\omega_p$ it holds that
\[\left.\deriv{}{\tau}\right|_{\tau = s} \beta_{\tau}^*(\omega_{\beta_\tau(p)}) = \alpha'(s)\omega_p.\]
Since $\mathcal L_T \omega_{\beta_s(p)} = -\theta(\beta_s(p))\omega_{\beta_s(p)}$ by \autoref{Lie derivative computation}, it holds that
\[\beta_s^*(\mathcal L_T \omega_{\beta_s(p)}) = -\theta(\beta_s(p)) \beta_s^*(\omega_{\beta_s(p)}) = -\theta(\beta_s(p)) \alpha(s) \omega_p.\]
Hence
\[\alpha'(s) = - \theta(\beta_s(p)) \alpha(s).\]
Solving this differential equation subject to the initial condition $\alpha(0) = 1$ we see that
\[\alpha(t) = \exp\left(-\int_0^t \theta(\beta_s(p))\,ds\right).\]
Since $\alpha(t) = J(\beta_t)(p)$ this completes the proof.
\end{proof}
\end{lem}

\subsubsection{Geodesically spanned null hypersurfaces}
\begin{prop}\label{Null geodesically spanned hypersurfaces are null}
Let $(M, g)$ be a spacetime of dimension $n+1$ and let $N \subset M$ be a spacelike $C^2$ submanifold of codimension $2$. Let $\mathbf n$ denote a $C^1$ normal null vector field along $N$. Consider the normal exponential map $\exp \colon \R \times N \to M$ defined by
\[\exp(t, p) = \exp_p(t \mathbf n_p)\]
where $\exp_p$ is the exponential map at the point $p$. Suppose that $\mathcal O \subset \R \times N$ is an open subset such that $\mathscr H := \exp(\mathcal O)$ is an embedded $C^1$ hypersurface in $M$ and the tangent map $\exp_*$ is injective on $\mathcal O$. Then $\mathscr H$ is a null hypersurface.
\begin{proof}
Choose a point $q = \exp(t, p) \in \mathscr H$ and let $\gamma$ denote the null geodesic $s \mapsto \exp(s, p)$. Our goal is to show that every vector $W \in T_q\mathscr H$ is orthogonal to $\dot \gamma(t)$, thereby proving that $T_q\mathscr H$ is a null hyperplane.

Since $\exp_* \colon T(\R \times N) \to T\mathscr H$ is injective at $(t, p)$, it is also surjective for dimensional reasons. This means that $W$ has some preimage in $T_{(t, p)}\left(\R \times N\right)$. Denote this preimage by $(\zeta, Z)$, where we make use of the canonical isomorphism $T_{(t, p)}\left(\R \times N\right) \cong T_t\R \times T_pN$. The pushforward is linear so
\[\exp_*(\zeta, Z) = \exp_*(\zeta, 0) + \exp_*(0, Z).\]
Note that $\exp_*(\zeta, 0)$ is tangent to the null curve $\gamma$, so $g(\exp_*(\zeta, 0), \dot \gamma(t)) = 0$. Hence
\[g(W, \dot \gamma(t)) = g(\exp_*(\zeta, 0) + \exp_*(0, Z), \dot \gamma(t)) = g(\exp_*(0, Z), \dot \gamma(t)).\]
Let $\alpha \colon (-1, 1) \to N$ be a curve with $\alpha(0) = p$ and $\dot\alpha(0) = Z$. Consider the two-parameter map
\[\mathbf x(s, u) = \exp(st, \alpha(u))\]
defined for $s \in [0, 1]$ and $u \in (-1, 1)$. Let $V$ be a vector field along $\gamma$ defined by
\[V(s) = \mathbf x_u(s, 0).\]
Each curve $s \mapsto \mathbf x(s, u)$ is a geodesic, so the map $\mathbf x$ is a variation through geodesics. Hence $V$ is a Jacobi vector field. The curve $u \mapsto \mathbf x(0, u)$ is contained in $N$ so $V(0)$ is tangent to $N$. By assumption on $\mathbf n$, the vector $\dot \gamma(0)$ is orthogonal to $N$, so
\[g(V(0), \dot \gamma(0)) = 0.\]
Let $T$ denote the vector field $\mathbf x_s$ along the map $\mathbf x$. Partial derivatives of two-parameter maps commute by \cite[Proposition 44, Chapter 4]{ONeill} so
\[V'(0) = \mathbf x_{us}(0, 0) = \mathbf x_{su}(0, 0) = \nabla_Z T.\]
Hence
\[g(V'(0), T) = g(\nabla_Z T, T) = \frac{1}{2}Zg(T, T) = 0\]
since $T$ is tangent to null curves. Since $\mathbf x_s(0, 0) = \dot \gamma(0)$ we have shown that
\[g(V'(0), \dot \gamma(0)) = 0.\]
By \cite[Lemma~7, Chapter 8]{ONeill}, the fact that $V(0)$ and $V'(0)$ are both orthogonal to the geodesic $\gamma$, together with the fact that $V$ is a Jacobi field along $\gamma$, implies that $V(s)$ is orthogonal to $\gamma$ for all $s$. In particular,
\[g(V(1), \dot \gamma(t)) = 0.\]
Computing $V(1)$ we see that
\[V(1) = \mathbf x_u(1, 0) = \exp_*(0, \dot \alpha(0)) = \exp_*(0, Z).\]
Hence
\[g(W, \dot \gamma(t)) = 0\]
for all $W \in T_q\mathscr H$. Since $q = \exp(t, p)$ was arbitrary, this shows that each tangent plane of $\mathscr H$ is a null hyperplane, so that $\mathscr H$ is a null hypersurface.
\end{proof}
\end{prop}

\subsection{Complete generators}
The following is a straightforward generalization of Lemma~8.5.5 in \cite{HawkingEllis}, and the proof follows that of \cite{HawkingEllis} but contains significantly more details. 
\begin{lem}\label{H+complete}
Let $S$ be an acausal set with $\edge(S) = \emptyset$ in a spacetime $(M,g)$ of dimension $n+1$. Let $\gamma$ be a null geodesic segment contained in $H^+(S)$. Suppose that $\gamma$ has no past endpoint and is totally past imprisoned in some compact set $K$. Suppose moreover that each point $p \in \im \gamma$ has some spacetime neighborhood $U_p$ such that $U_p \cap \im \gamma$ is contained in a $C^{1,1}$ hypersurface $N_p$.
Then $\gamma$ is complete in the past direction. 
\begin{proof}
Let $\gamma$ have an affine parametrization.
Suppose to get a contradiction that $\gamma$ is incomplete to the past, i.e. that the domain of $\gamma$ has some infimum $v_0$. We may without loss of generality (by translation of the parameter of $\gamma$ and restriction of $\gamma$ to a smaller domain to the future) assume that $\gamma$ has domain $(v_0, 0]$ and that $\gamma(t) \in K$ for all $t \in (v_0, 0]$. Then the set $\overline{\im \gamma}$ is compact, so we may assume without loss of generality that $K = \overline{\im \gamma}$. Let $\mathcal W$ be a neighborhood of $K \cap H^+(S)$ with compact closure.

The idea is now to show that if $\gamma$ is past incomplete, then a small perturbation of it yields a past inextendible timelike curve with contradictory properties. To help with this, we will introduce a timelike vector field $V$. For the construction of $V$, we will need an auxiliary distance function compatible with the manifold topology, for instance one given by a Riemannian metric $\eta$. Fix such a distance function and call it $d_\eta$.
Since $M$ is time-orientable, it admits a future-directed timelike vector field. Fix such a vector field and call it $Z$. For each $p \in \im \gamma$, we will define a vector field $V^p$ in a neighborhood of $p$ with the following properties.
\begin{itemize}
\item $V^p$ is timelike and future-directed.
\item $V^p = Z$ on $\im \gamma \cap \operatorname{dom}(V^p)$.
\item $\nabla_{V^p} V^p = 0$ on all integral curves passing through $\im \gamma$.
\end{itemize}
To do this, consider a small neighborhood of $p$ whose intersection with $\gamma$ is contained in a $C^{1,1}$ hypersurface $N$. Such a neighborhood exists by hypothesis. Without loss of generality we may assume that $T_qN$ does not contain any timelike vectors, since $\im \gamma \subseteq H^+(S)$ and hence cannot accumulate on itself in a timelike direction. Consider the restriction of the exponential map to the restriction to $N$ of the subbundle of $TM$ spanned by $Z$. In other words, consider the map $\exp_{Z} \colon N \times \R \to M$ defined by
\[\exp_{Z}(q, t) = exp_q(t Z).\]
This map is submersive at $(p, 0)$, and hence for dimensional reasons also immersive at $(p, 0)$. By the inverse function theorem, it is then a $C^{1,1}$ diffeomorphism on some open neighborhood $(p, 0)$. Let $\mathcal U_p$ be the image of this neighborhood under $\exp_{Z}$. Let $\rho(p) > 0$ be a real number such that all points $r \in M$ with $d_\eta(r, p) < 4\rho(p)$ belong to $\mathcal U_p$. (Note that we will not need any continuity of $\rho$.) Let $\mathcal W_p$ be the set of points $r \in M$ with $d_\eta(r, p) < \rho(p)$. Define $V^p$ on $\mathcal W_p$ to be the tangent vectors of the curves $s \mapsto \exp_{Z}(q, s)$. Since $\exp_{Z}$ is a diffeomorphism onto $\mathcal W_p$, this is well-defined.

We will now show that if $r \in \mathcal W_p \cap \mathcal W_q$ for some $p, q \in \im \gamma$ is such that the integral curves of both $V^p$ and $V^q$ through $r$ both intersect $\gamma$, then $V^p_r = V^q_r$. Suppose that $r \in \mathcal W_p \cap \mathcal W_q$, that the integral curve of $V^p$ through $r$ intersects $\im\gamma$ in $r_p$, and that the integral curve of $V^q$ through $r$ intersects $\im\gamma$ in $r_q$. Suppose without loss of generality that $\rho(p) \geq \rho(q)$. Then $d_\eta(p, r_q) \leq d_\eta(p, r) + d_\eta(r, q) + d_\eta(q, r_q) < \rho(p) + 2\rho(q) \leq 3\rho(p)$. Hence $r_q \in \mathcal U_p$. Since $V^p_{r_q} = V^q_{r_q} = {Z}_{r_q}$ and $\exp_{Z}$ is a diffeomorphism onto $\mathcal U_p$, this means that $r_q = r_p$ and, by following the geodesic from $r_q = r_p$ with initial velocity $V^p_{r_q} = V^q_{r_q} = {Z}_{r_q}$, that $V^p_r = V^q_r$.

Choose a countable subset $C$ of $\im \gamma$ such that the sets $\{\mathcal W_p\}_{p \in C}$ cover $\im \gamma$. Combine the vector fields $V^p$ for $p \in C$ using a partition of unity corresponding to this cover to obtain a vector field $V$. This vector field is Lipschitz, timelike and future-directed since the $V^p$ are. Moreover, $\nabla_V V = 0$ on each integral curve of $V$ passing through $\im \gamma$, since this holds for the $V^p$ and they agree on such curves. By a further partition of unity, we may extend $V$ to a future-directed timelike Lipschitz vector field on all of $M$. Note, however, that $V$ has larger regularity than Lipschitz on a $2$-dimensional surface close to $\im \gamma$. More precisely, there is a subset $\Omega = \{(t, u) \in \operatorname{dom}(\gamma)\times\R \mid |u| < \psi(t)\}$ for some positive smooth function $\psi$ such that the map $\Omega \to TM$ defined by
\[(t, u) \mapsto V_{\exp_{\gamma(t)}(u Z_{\gamma(t)})}\]
is smooth. However, $V$ is not necessarily smooth when viewed as a vector field on the spacetime.

Define a metric $g'$ by
\[g'(X, Y) = g(X, Y) + 2g(X, V)g(Y, V).\]
This metric is positive definite. To see this, let $X$ be nonzero and compute $g'(X, X)$ in a basis $(V, e_1, e_2, \ldots, e_n)$ orthonormal for $g$:
\[\begin{split}
g'(X, X) &= g(X, X) + 2(g(X, V))^2\\
	 &= \left(- (X^0)^2 + (X^1)^2 + (X^2)^2 + \cdots + (X^n)^2\right) + 2(X^0)^2 > 0.
\end{split}\]

Let $\alpha_0(t) = \gamma(v(t))$ be a reparametrization of $\gamma$ such that $g(\dot{\alpha_0}, V) = -1/\sqrt{2}$.
Note that $v$ is a smooth function, since $V$ is smooth when viewed as a vector field along $\gamma$. This means that $\alpha_0$ is a smooth curve.
The definition of $v$ implies that $v$ is strictly increasing. For convenience, suppose also that $v(0) = 0$.
Note that $\alpha_0$ is parameterized by arc length in the Riemannian metric $g'$:
\[\begin{split}
\int_a^b \sqrt{g'(\dot\alpha_0(t), \dot \alpha_0(t))}\,dt &= \int_a^b \sqrt{g(\dot \alpha_0(t), \dot \alpha_0(t)) + 2(g(\alpha_0(t), V))^2}\,dt\\
&= \int_a^b \sqrt{0 + 2 \frac{1}{2}}\,dt = b-a
\end{split}\]
Since $\gamma$ has no past endpoint, $\alpha_0$ does not have one either.

We will later construct a variation $\alpha$ of $\alpha_0$, and the computations will be done along the two-parameter map $\alpha$.

\begin{claim}{The domain of $\alpha_0$ is not bounded from below}
Suppose for contradiction that the domain of $\alpha_0$ were bounded below. Let $a > -\infty$ be the infimum of the domain of $\alpha_0$.
Recall that a Riemannian metric induces a distance function defined as the infimum of the lengths of curves from one point to another. 
Then for any sequence $a_n \to a$ it holds that $\alpha_0(a_n)$ is a Cauchy sequence with respect to the distance function induced by $g'$ (for $\alpha_0$ is a curve of length $|a_n - a_m| < |\max(a_m, a_n) - a| \to 0$ connecting $\alpha_0(a_n)$ to $\alpha_0(a_m)$). The sequence $\alpha_0(a_n)$ is also contained in the compact set $K$, and so has a convergent subsequence. These two statements together imply that $\alpha_0(a_n)$ is convergent for any sequence $a_n \to a$ so the limit $\lim_{t \to a^+} \alpha_0(t)$ exists contradicting the fact that $\alpha_0$ has no past endpoint. Hence the domain of $\alpha_0$ is not bounded from below.
\end{claim}

\begin{proofpart}{Relations between $\alpha_0$ and $\gamma$}
Since $\alpha_0$ is a reparametrization of a geodesic, $\nabla_{\dot \alpha_0} \dot \alpha_0$ is parallel to $\dot \alpha_0$. In other words, there is a function $f \colon (-\infty, 0) \to \R$ such that
\[\nabla_{\dot \alpha_0(t)} \dot \alpha_0(t) = f(t) \dot \alpha_0(t), \quad \forall t \in (-\infty, 0).\]
Note that $f$ is a smooth function.
It also holds that
\[v'(t) \dot \gamma(v(t)) = \dot \alpha_0(t), \quad \forall t \in (-\infty, 0).\]
Now
\[f(t) \dot \alpha_0(t) = \nabla_{\dot \alpha_0} \dot \alpha_0 = \nabla_{\dot \alpha_0} (v' \dot \gamma) = \alpha_0(v')\dot\gamma + v'\nabla_{\dot \gamma} \dot \gamma = \frac{v''(t)}{v'(t)} \dot\alpha_0(t)\]
so
\[f = \frac{v''}{v'}.\]

Note also that $f$ is bounded. This can be seen by the following computation.
\[f = -\sqrt{2} g(f \dot \alpha_0, V)
= -\sqrt{2} g(\nabla_{\dot \alpha_0} \dot \alpha_0, V)
= -\sqrt{2}\left(\nabla_{\dot \alpha_0} g(\dot \alpha_0, V) - g(\dot \alpha_0, \nabla_{\dot \alpha_0} V)\right)\]
\[= -\sqrt{2}\left(\dot \alpha_0(-1/\sqrt{2}) - g(\dot \alpha_0, \nabla_{\dot \alpha_0} V)\right)
= \sqrt{2} g(\dot \alpha_0, \nabla_{\dot \alpha_0} V).\]
This shows that $f$ can be defined in terms of $g$, $\dot \alpha$ and $V$. The coordinate representations of these objects in coordinate patches are all bounded since $\dot \alpha$ is a unit vector field in $g'$ and $V$ is a Lipschitz continuous vector field which is smooth along $\alpha_0$. 
Since $H^+(S) \cap K$ is compact it can be covered by finitely many coordinate patches, and hence $f$ is bounded.
\end{proofpart}

\begin{claim}{$v'$ is bounded}
Since $\gamma$ is incomplete to the past, $v$ is bounded below. In other words, the integral
\[v(t) = \int_0^t v'(\tau) d\tau\]
is bounded. This implies that $\liminf_{t \to -\infty} v'(t) = 0$, since $v$ is strictly increasing. We will now show that boundedness of $v$ on $(-\infty, 0]$ together with boundedness of $f = \frac{v''}{v'}$ implies that $v'$ is bounded. Suppose not. Since $v'$ is continuous, it can only be unbounded on $(-\infty, 0]$ if $\limsup_{t \to -\infty} v'(t) = \infty$. Since we also know that $\liminf_{t \to -\infty} v'(t) = 0$ and that $v'$ is continuous there are, for arbitrarily large $C > 0$, sequences $t_n, s_n \to -\infty$ such that
\[t_{n+1} < s_n < t_n \text{ for all $n$},\]
\[v'(t_n) = 2C,\]
\[v'(s_n) = C\]
and
\[C \leq v'(t) \leq 2C \text{ if $t \in (s_n, t_n)$.}\]
By the mean value theorem of calculus, there is for each $n$ some $\tau_n \in [s_n, t_n]$ such that
\[v''(\tau_n) = \frac{v'(t_n) - v'(s_n)}{t_n - s_n} = \frac{C}{t_n - s_n}.\]
However
\[\sum_{n = 0}^\infty C(t_n - s_n) \leq \left|\int_0^{-\infty} v'(\tau) d\tau\right| < \infty\]
so $(t_n - s_n) \to 0$ as $n \to \infty$. Hence
\[\lim_{n \to \infty} f(\tau_n) v'(\tau_n) = \lim_{n \to \infty} v''(\tau_n) = \infty.\]
Since $v'(\tau_n) \in [C, 2C]$ for all $n$, this implies that $f(\tau_n) \to \infty$, contradicting the fact that $f$ is bounded. Hence $v'$ must be bounded.
\end{claim}

\begin{proofpart}{Construction of a variation $\alpha$ of $\alpha_0$}
We will now construct a variation $\alpha$ of $\alpha_0$. The idea is to push $\alpha_0$ to the past and make it timelike, and then derive a contradiction from the resulting curve. 
Let $x \colon (-\infty, 0) \to \R$ denote a smooth positive function which will be fixed later. Let
\[\begin{aligned}
\alpha \colon (-\delta, \delta) \times (-\infty, 0)	&\to H^+(S)\\
					(u, t)		&\mapsto \alpha(u, t)
\end{aligned}\]
be a smooth map such that
\begin{equation}\label{Definition of the variation alpha}
\alpha(0, \cdot) = \alpha_0\ \text{ and }\ \partderiv{\alpha}{u}(u, t) = -x(t) V_{\alpha(u, t)}.
\end{equation}
Recall that $V_{\alpha(u, t)}$ is smooth as a function of $u$ and $t$, even though $V$ is not a smooth vector field on the spacetime.
To see that such a variation exists, note that the conditions can be viewed as a family of ordinary differential equations in $u$, parameterized by $t$. As a consequence of the existence theorem and theorem about smooth dependence on initial values for ordinary differential equations there is, for each $t$, a smooth solution with existence time $\delta_t > 0$. To claim that the necessary variation exists, we need to show that the existence times $\delta_t$ can be uniformly bounded from below by some $\delta > 0$ independent of $t$. However, we know that a solution to the differential equation exists as long as it stays in the compact set $\overline{\mathcal W}$. Since $H^+(S) \cap K$ is compact and $\mathcal W$ open, the $g'$ distance between $H^+(S) \cap K$ and $M \setminus \mathcal W$ is positive. Since $V$ is bounded, and $x$ will be bounded when we choose it, there is a positive uniform lower bound for the time after which a solution may leave $\mathcal W$. This means that there is a uniform lower bound for the existence times of the solutions of the family of ordinary differential equations defining the variation. Hence we may choose a suitable $\delta > 0$ uniformly, and a variation with the desired properties exists.

Let $\alpha_u$ denote the curve $\alpha(u, \cdot)$. Note that each curve $\alpha_u$ is smooth. We now wish to choose the positive function $x$ in such a way that some curve $\alpha_\epsilon$ is timelike. In other words, we want there to be some $\epsilon > 0$ such that the function
\[y(u, t) = g(\dot \alpha_u(t), \dot \alpha_u(t))\]
is negative for $u = \epsilon$ and all $t \in (-\infty, 0)$. To show that this is the case, we will compute $\left.\frac{\partial y}{\partial u}\right|_{u = 0}$ and a bound for $\frac{\partial^2 y}{\partial u^2}$, and from this obtain an upper bound for $y$. Choosing a suitable function $x$ will make this upper bound negative for small values of $u$.
\end{proofpart}

\begin{proofpart}{Computation of $\displaystyle \partderiv{y}{u}$}
Let $U$ denote the pushforward through $\alpha$ of the coordinate vector field $\partderiv{}{u}$ on $(-\delta, \delta) \times (-\infty, 0)$. We will not always write out the dependence on $t$ and $u$.
The first partial derivative of $y$ can be computed as
\[\frac{\partial y}{\partial u}(u, t)
= \frac{\partial}{\partial u} g\left(\dot \alpha_u(t), \dot \alpha_u(t)\right)
= U g(\dot \alpha_u, \dot \alpha_u)
= 2 g(\nabla_{U} \dot \alpha_u, \dot \alpha_u)\]
\[= 2 g(\nabla_{\dot \alpha_u} U, \dot \alpha_u)
= 2\left(\nabla_{\dot \alpha_u} g(U, \dot \alpha_u) - g(U, \nabla_{\dot \alpha_u}\dot \alpha_u)\right)\]
where $\nabla_{U} \dot \alpha_u = \nabla_{\dot \alpha_u} U$ since $U$ and $\dot\alpha_u$ are pushforwards of coordinate vector fields.
Evaluating this at $u = 0$ we see that
\[\begin{split}
\frac{\partial y}{\partial u}(0, t)
&= 2\left(\nabla_{\dot \alpha_0} g(-x V, \dot \alpha_0) - g(-x V, \nabla_{\dot \alpha_0}\dot \alpha_0)\right)\\
&= 2\left(-\nabla_{\dot \alpha_0} (x g(V, \dot \alpha_0)) + x g(V, \nabla_{v' \dot \gamma}(v' \dot \gamma))\right)\\
&= 2\left(\frac{1}{\sqrt{2}} \dot \alpha_0 (x) + x v' g(V, \nabla_{\dot \gamma}(v' \dot \gamma))\right)\\
&= 2\left(\frac{1}{\sqrt{2}} \dot\alpha_0(x) + x (v')^2 g(V, \nabla_{\dot \gamma}(\dot \gamma)) + \frac{x}{v'}\dot \alpha_0(v') g(V, \dot \alpha_0)\right)\\
&= \sqrt{2} x'(t) - \sqrt{2}\frac{x(t) v''(t)}{v'(t)}\\
&= \sqrt{2} v'(t) \deriv{}{t} \left(\frac{x(t)}{v'(t)}\right),
\end{split}\]
where we have used that $\dot \alpha_0 = v'\dot \gamma$, $g(V, \dot \alpha_0) = - 1/\sqrt{2}$ and $\nabla_{\dot \gamma} \dot \gamma = 0$.
\end{proofpart}

\begin{proofpart}{An upper bound for $\displaystyle \partdderiv{y}{u}$}
We now compute an upper bound for the second partial derivative of $y$ with respect to $u$. 
For convenient notation, we use the vector fields
\[T = \alpha^*\left(\partderiv{}{t}\right),\]
\[U = \alpha^*\left(\partderiv{}{u}\right).\]
Note that
\[T_{\alpha(u, t)} = \dot \alpha_u(t)\]
and
\[U_{\alpha(u, t)} = -x(t) V_{\alpha(u, t)}.\]
Now
\[\frac{1}{2}\frac{\partial^2}{\partial u^2} y(u, t)
= \frac{1}{2}\frac{\partial^2}{\partial u^2} g\left(\dot \alpha_u(t), \dot \alpha_u(t)\right)
= \frac{1}{2}\frac{\partial^2}{\partial u^2} g\left(T, T\right)
= \partderiv{}{u} g(\nabla_U T, T)\]
\[= g(\nabla_U T, \nabla_U T) + g(\nabla_U \nabla_U T, T) 
= g(\nabla_T U, \nabla_T U) + g(\nabla_U \nabla_T U, T)\]
\[= g(\nabla_T U, \nabla_T U) + g(\nabla_T \nabla_U U, T) + g(R(U, T)U, T)\]
where we have used that $\nabla_U T = \nabla_T U$ since $U$ and $T$ are coordinate vector fields and
\[\nabla_{U}\nabla_{T} = \nabla_T\nabla_{U} + R(U, T).\]
We now compute each term separately. 

Evaluating the first term at $\alpha(0, t)$ and using that $T(x) = \alpha_0(x) = x'$ we get
\[g(\nabla_T U, \nabla_T U) = g(\nabla_T (xV), \nabla_T (xV)) = g(T(x)V + x \nabla_T V, T(x)V + x \nabla_T V)\]
\[= \left(x'(t)\right)^2 g(V, V) + 2x(t)x'(t) g(V, \nabla_T V) + x^2(t)g(\nabla_T V, \nabla_T V)\]
\[= -\left(x'(t)\right)^2 + (x(t))^2 g(\nabla_T V, \nabla_T V).\]
We have used that $g(V, \nabla_T V) = 0$. That this is true is seen by noting that
\[g(V, \nabla_T V) = Tg(V, V) - g(\nabla_T V, V) = T(-2^{-1/2}) - g(V, \nabla_T V) = -g(V, \nabla_T V)\]
so that $g(V, \nabla_T V) = -g(V, \nabla_T V)$.
For the second term, note that
\[\nabla_U U = \nabla_U (-x V) = xV(x) V + x^2 \nabla_V V = - \partderiv{x}{u}V + 0 = 0\]
(since $\nabla_V V = 0$ on the image of $\alpha$ by choice of $V$, and $x$ is independent of $u$) so that
\[g(\nabla_T\nabla_U U, T) = g(\nabla_T 0, T) = 0.\]
The third term is simply
\[g(R(U, T) U, T) = g(R(-xV, T)(-xV), T) = x^2(t)g(R(V, T)V, T).\]
Hence
\[\frac{1}{2}\frac{\partial^2}{\partial u^2} g\left(\dot \alpha_u(t), \dot \alpha_u(t)\right)
= -\left(x'(t)\right)^2 + (x(t))^2 \left(g(\nabla_T V, \nabla_T V) + g(R(V, T)V, T)\right)\]
\[\leq x^2 \left(g(\nabla_T V, \nabla_T V) + g(R(V, T)V, T)\right).\]
We wish to bound this by $C^2 x^2 g'(T, T)$ for some constant $C$ on the neighborhood $\mathscr{W}$ of $H^+(S)$, which we chose to have compact closure. (Recall that $g'$ is the Riemannian metric constructed from the vector field $V$ in the beginning of the proof.) To see that this is possible, view $g(\nabla_T V, \nabla_T V) + g(R(V, T)V, T)$ as a quadratic form in $T$. Its components in coordinates depend on $g$, $R$, $V$ and derivatives of $V$, all of which are bounded in coordinate neighborhoods since $V$ is a Lipschitz continuous vector field, and $H^+(S) \cap K$ can be covered by finitely many such neighborhoods. Since the quadratic form $g'$ is positive definite, there is some $C$ such that $g(\nabla_T V, \nabla_T V) + g(R(V, T)V, T) \leq Cg'(T, T)$. Hence
\[\partdderiv{}{u} g\left(\dot \alpha_u(t), \dot \alpha_u(t)\right)
\leq C^2 x^2 g'(T, T)\]
for some constant $C$. We want a bound in terms of $g\left(\dot \alpha_u(t), \dot \alpha_u(t)\right)$ instead, so we compute
\[g'(T, T) = g(T, T) + 2\left(g(V, T)\right)^2.\]
Since
\[\partderiv{}{u} g(V, T)
= U g(V, T)
= g(-x \nabla_V V, T) + g(V, \nabla_U T)
= 0 + g(V, \nabla_T U)\]
\[= -g(V, T(x)V - x\nabla_T V)
= T(x) + x g(V, \nabla_T V)
= x'(t)\]
(where as earlier $g(V, \nabla_T V) = 0$) we know that
\[g(V, T) = u x'(t) + \left.g(V, T)\right|_{u = 0} = u x'(t) - \frac{1}{\sqrt{2}}.\]
When we choose $x$, we will make sure that $\deriv{x}{t}$ is bounded, and then $2\left(g(V, T)\right)^2$ is bounded by some constant $d$ for all small $u$. 
Hence we can convert our bound in terms of $g'(T, T)$ to a bound in terms of $g(T, T)$:
\[\partdderiv{}{u} g\left(\dot \alpha_u(t), \dot \alpha_u(t)\right)
\leq C^2 x^2 g'(T, T) \leq C^2x^2 (g(T, T) + d).\]
In the notation of the function $y$, we now know that
\[\frac{\partial^2 y}{\partial u^2}(u, t) \leq (y(u, t) + d)C^2(x(t))^2\]
for all sufficiently small $u > 0$.
\end{proofpart}

\begin{claim}{For all sufficiently small $\epsilon > 0$, the curve $\alpha_\epsilon$ is timelike}
From our previous computations we know that
\[\frac{\partial y}{\partial u}(0, t) = \frac{v'(t)}{\sqrt{2}} \deriv{}{t} \left(\frac{x(t)}{v'(t)}\right),\]
\[\frac{\partial^2 y}{\partial u^2}(u, t) \leq (y(u, t) + d)C^2(x(t))^2.\]
Moreover, $y(0, t) = 0$ since $\alpha_0$ is a lightlike curve. For each fixed $t$, this is a differential inequality in the variable $u$.
Let $z$ be the solution of the differential equation resulting from replacing the inequality with equality:
\[\frac{\partial^2 z}{\partial u^2}(u, t) = C^2x^2(0, t)(z(u, t) + d),\]
\[\frac{\partial z}{\partial u}(0, t) = \frac{\partial y}{\partial u}(0, t),\]
\[z(0, t) = y(0, t) = 0.\]
Integrating the inequality $\frac{\partial^2 y}{\partial u^2}(u, t) \leq \frac{\partial^2 z}{\partial u^2}(u, t)$ we see that
\[\frac{\partial y}{\partial u}(u, t) - \frac{\partial y}{\partial u}(0, t) \leq \frac{\partial z}{\partial u}(u, t) - \frac{\partial z}{\partial u}(0, t)\]
so that
\[\frac{\partial y}{\partial u}(u, t) \leq \frac{\partial z}{\partial u}(u, t).\]
Integrating once again and using the fact that $z(0, t) = y(0, t) = 0$ we have
\[y(u, t) \leq z(u, t).\]
Solving the differential equation for $z$ we see that
\[z(u, t) = d\cosh(C x(t) u) + a(t) \sinh(C x(t) u) - d\]
where
\[a(t) = \frac{v'(t)}{\sqrt{2}C x(t)} \deriv{}{t}\left(\frac{x(t)}{v'(t)}\right).\]
Since $d$ is nonnegative, an upper bound for $z$ is
\[\begin{split}
z(u, t) &= d\cosh(C x(t) u) + a(t) \sinh(C x(t) u) - d\\
	&= (d \tanh(C x(t) u) + a(t))\sinh(C x(t) u) - d\\
	&\leq (d \tanh(C x(t) u) + a(t))\sinh(C x(t) u).
\end{split}\]
Hence
\[y(u, t) \leq (d \tanh(C x(t) u) + a(t))\sinh(C x(t) u).\]

Recall that the idea was to choose the function $x$ in such a way that there exists some $\epsilon > 0$ such that $y(\epsilon, t) < 0$ for all $t$. We claim that an example of such a function $x$ is
\[x(t) = \frac{v'(t)}{v(t) - 2v_0}.\]
Recall that
\[v_0 = \lim_{t \to -\infty} v(t)\]
and that $v$ is increasing so that
\[v_0 \leq v(t) \leq 0 \quad \forall t \in (-\infty, 0].\]
We begin by making good on the promises we made about the function $x$: It should be positive, bounded, and have bounded derivative. The denominator in the definition of $x$ is bounded from below by $-v_0$ and from above by $-2v_0$, and $-v_0$ is positive, so boundedness and positivity of $x$ follow from boundedness and positivity of $v'$. Computing the derivative of $x$ we see that
\[x'(t) = \frac{v''(t)}{v(t) - 2v_0} - \frac{(v'(t))^2}{(v(t) - 2v_0)^2} = \frac{v'(t)f(t)}{v(t) - 2v_0} - x^2(t)\]
Since $x$, $v'$ and $f = v''/v'$ are bounded, so is $x'$. Having chosen $x$, we can now fix the number $\delta > 0$ defining the domain of $\alpha$ such that the image of $\alpha$ is contained in $\mathscr W$.

Recall that
\[y(u, t) \leq (d \tanh(C x(t) u) + a(t))\sinh(C x(t) u)\]
where
\[a(t) = \frac{v'(t)}{\sqrt{2}C x(t)} \deriv{}{t}\left(\frac{x(t)}{v'(t)}\right).\]
With our present choice of $x$,
\[a(t) = -\frac{x(t)}{\sqrt{2}C v'(t)}.\]
The objective is to ensure that $y(u, t) < 0$ for some positive $u$ and for all $t$. Since $\sinh(Cxu) \geq 0$ for positive $u$, a sufficient condition is that
\[d \tanh(C x(t) u) - \frac{x(t)}{\sqrt{2}C v'(t)} < 0\]
for some $u > 0$ and all $t$. A series expansion tells us that
\[\tanh(C x(t) u) = C x(t) u + \mathcal{O}((ux(t))^3)\]
for small $ux(t)$, so that
\[d \tanh(C x(t) u) + a(t) = \left(dCu - \frac{1}{\sqrt{2}C v'(t)}\right)x(t) + \mathcal O(u^3x^3(t)).\]
Since $v'$ is bounded, there is some positive lower bound for $1/v'$. Hence it holds for all sufficiently small $u$ such that $dCu - 1/(\sqrt{2}C v'(t))$ is negative for all $t$. Since $x$ is bounded, it further holds for all sufficiently small $u$ that the $\mathcal O(u^3x^3(t))$ term does not affect the sign: With such a choice of $u$, it holds that $d \tanh(Cxu) + a$ is negative for all $t$, and hence
\[y(\epsilon, t) \leq (d \tanh(Cx(t)\epsilon) + a(t))\sinh(Cx(t)\epsilon) < 0\]
for all values of $t$ and all sufficiently small $\epsilon > 0$.
Since
\[y(\epsilon, t) = g(\dot \alpha_\epsilon(t), \dot \alpha_\epsilon(t))\]
this shows that the curve $\alpha_\epsilon$ is timelike.
\end{claim}

\begin{claim}{For all sufficiently small $\epsilon > 0$, the curve $\alpha_\epsilon$ has infinite $g'$-length in the past direction}
For each (negative) integer $k$, let $L_k(u)$ be the $g'$-length of the restriction of $\alpha_u$ to $[k, k+1]$. By the formula for the first variation of arc length (\cite[Proposition 2, Chapter 10]{ONeill})
\[\begin{split}
L_k'(0) &= -\int_k^{k+1} g'(\nabla_{\dot \alpha_0} \dot \alpha_0, V)\,dt + \left.g'(\dot \alpha_0, V)\right|_k^{k+1}\\
&= -\int_k^{k+1} f(t)g'(\dot \alpha_0, V)\,dt + \left.g'(\dot \alpha_0, V)\right|_k^{k+1}\\
&= -\int_k^{k+1} \frac{f(t)}{\sqrt{2}}\,dt.
\end{split}\]
We here used that $g'(\alpha_0, V) = 1/\sqrt{2}$ by definition of $g'$ and $\alpha_0$. Since $f$ is bounded, we know that $L_k'(0)$ is bounded uniformly in $k$. This means that for all sufficiently small $\epsilon > 0$ it holds that $L_k(\epsilon) > 1/2$ for all $k$. (Recall that $L_k(0) = 1$ since $\alpha_0$ is parameterized by arc length.) This means that the length of $\alpha_\epsilon$ is
\[\sum_{k < 0} L_k(\epsilon) \geq \sum_{k < 0} 1/2 = \infty.\]
\end{claim}

\begin{claim}{For all sufficiently small $\epsilon > 0$, the curve $\alpha_\epsilon$ belongs to the interior of $D^+(S)$}
Since the curve $\alpha_\epsilon$ for $\epsilon > 0$ is a variation to the past of $\alpha_0$, it belongs to the open set $I^-(H^+(S))$. We will first show that $I^-(H^+(S)) \cap I^+(S) \subseteq D^+(S)$, and then show that $\alpha_\epsilon$ belongs to $I^-(H^+(S)) \cap I^+(S)$ for all sufficiently small $\epsilon > 0$. We will then have shown that $\alpha_\epsilon$ belongs to an open set contained in $D^+(S)$, and hence it must belong to the interior of $D^+(S)$.

Let $p \in I^-(H^+(S)) \cap I^+(S)$. We will first show that $p \in \overline{D^+(S)}$, and then that $p \in D^+(S)$. That $p \in I^-(H^+(S))$ means that there is some future-directed timelike curve $\lambda$ from $p$ to $H^+(S)$. This curve cannot pass $S$, since $p$ lies to the future of $S$ and $S$ is acausal. Suppose now that $\kappa$ is a future-directed past inextendible timelike curve with future endpoint $p$. By concatenating $\kappa$ and $\lambda$ and smoothing (in a neighborhood of $p$ which is disjoint from $S$, which exists since $p \in I^+(S)$) we obtain a past-inextendible timelike curve with future endpoint in $H^+(S)$. Since $H^+(S) \subseteq \overline{D^+(S)}$, this combined curve must intersect $S$. Since the curve $\lambda$ does not intersect $S$, the curve $\kappa$ must do so. This proves that every past-inextendible timelike curve $\kappa$ through $p$ must intersect $S$, so that $p \in \overline{D^+(S)}$. By the same argument, all points in the interior of $\lambda$ belong to $\overline{D^+(S)}$. Let $q$ be some point in the interior of $\lambda$. Since $\lambda$ is timelike, $q \in I^+(p)$. Since $I^+(p)$ is open, it is a neighborhood of $q$. Since $q \in \overline{D^+(S)}$ it is a limit point of $D^+(S)$. This means that the neighborhood $I^+(p)$ of $q$ must contain some point $r \in D^+(S) \cap I^+(p)$. Let $\widehat \lambda$ be a future-directed timelike curve from $p$ to $r$. Now let $\kappa$ be a future-directed past inextendible causal curve with future endpoint $p$. Concatenating $\kappa$ with $\widehat \lambda$ and smoothing (again in a neighborhood of $p$ which is disjoint from $S$) we obtain a past-inextendible causal curve with future endpoint $r$. Since $r \in D^+(S)$, this curve must intersect $S$. Since $r \in I^+(p) \subseteq I^+(S)$ and $S$ is acausal, the curve $\widehat\lambda$ cannot intersect $S$. This means that $\kappa$ must intersect $S$. This proves that every past-inextendible causal curve $\kappa$ through $p$ must intersect $S$, so that $p \in D^+(S)$.

Since $\alpha_0$ is a curve in $H^+(S)$ and $\alpha_\epsilon$ is a variation to the past for $\epsilon > 0$ we know that $\alpha_\epsilon$ belongs to $I^-(H^+(S))$. 
By \cite[Corollary~26, Chapter 14]{ONeill}, the acausal edgeless set \(S\) is a closed topological hypersurface, and by \cite[Lemma~43, Chapter 14]{ONeill} it belongs to the interior of \(D^+(S)\). Since \(H^+(S) \subseteq \partial D^+(S)\) this means that \(S\) and \(H^+(S)\) are disjoint. Since \(S\) is closed and \(H^+(S) \cap K\) is compact, the \(g'\)-distance from \(H^+(S) \cap K\) to \(S\) is positive.
Since $g'(V, V) = 1$ and $x$ is bounded by $|2v_0|$, equation \eqref{Definition of the variation alpha} implies that the distance from a point on $\alpha_\epsilon$ to $\alpha_0$ cannot exceed $|2v_0\epsilon|$. Choosing $\epsilon > 0$ so small that $|2v_0\epsilon|$ is smaller than the \(g'\)-distance from \(H^+(S) \cap K\) to \(S\), we know that $\alpha_\epsilon$ does not intersect $S$. To see that $\alpha_\epsilon(t) \in I^+(S)$ for some $t$, note that no curve $\alpha_u$ with $0 \leq u \leq \epsilon$ can intersect $S$ so that the timelike curve $\lambda \colon [-\epsilon, 0] \to M$ defined by $\lambda(u) = \alpha_{-u}(t)$ does not intersect $S$. Extend $\lambda$ to some past inextendible timelike curve. Then $\lambda$ is a past inextendible timelike curve with future endpoint $\lambda(0) = \alpha_0(t) \in H^+(S)$, so $\lambda$ must intersect $S$. Since $\lambda$ passes through $\alpha_\epsilon(t)$, we know that $\alpha_\epsilon(t) \in I^+(S)$. Since $t$ was arbitrary, we have now shown that the image of $\alpha_\epsilon$ belongs to $I^-(H^+(S)) \cap I^+(S)$ for all sufficiently small $\epsilon > 0$. As noted previously, this together with the fact that $I^-(H^+(S)) \cap I^+(S) \subseteq D^+(S)$ shows that the image of $\alpha_\epsilon$ belongs to the interior of $D^+(S)$.
\end{claim}

\begin{proofpart}{Contradiction}
We have now shown that if we choose $\epsilon > 0$ small enough, then $\alpha_\epsilon$ is a timelike curve of infinite $g'$-length, contained in the interior of $D^+(S)$. 
Since it has infinite $g'$-length in the past direction, it is past inextendible. This means that we have constructed a past inextendible timelike curve passing through a point in the interior of $D^+(S)$ without intersecting $S$. This is a contradiction.

Hence $\gamma$ cannot be incomplete in the past direction.
\end{proofpart}
\end{proof}
\end{lem}

\subsection{Structure of horizons}\label{Structure of horizons}
We begin by defining the abstract concept of a ''horizon'' (following \cite{CDGH01}), and state some previously known results about the regularity of horizons. We then prove that the Cauchy horizons we will work with are horizons in this sense.

\subsubsection{Abstract horizons}\label{Abstract horizons}
\begin{defn}
We say that an embedded topological hypersurface in a spacetime is \emph{past null geodesically ruled} if every point on the hypersurface belongs to a past inextendible null geodesic contained in the hypersurface. These geodesics are called \emph{generators}.
\end{defn}

\begin{remark}
Note that if a past null geodesically ruled hypersurface is a $C^2$ null hypersurface, then these generators are the same as those defined in \autoref{The null Weingarten map}.
\end{remark}

\begin{defn}\label{Definition of horizon}
A \emph{horizon} in a spacetime is an embedded, achronal, past null geodesically ruled, closed (as a set) topological hypersurface.
\end{defn}
\begin{remark}
One may just as well define a horizon to be future null geodesically ruled. Indeed, in \cite{CDGH01} the distinction is made between a ''past horizon'' and a ''future horizon''. However, since we will work only with future Cauchy horizons it is convenient to restrict our attention to past null geodesically ruled horizons.
\end{remark}
\begin{remark}
If an open subset of a horizon is past null geodesically ruled, then its generators are the restrictions of the generators of the horizon.
\end{remark}
Note that we have assumed no smoothness in the definition. Note also that the generators through a point of a horizon need not be unique. In fact, we have the following theorem (see Theorem~3.5 in \cite{BeemKrolak98} and Proposition 3.4 in \cite{ChruscielGalloway98}).
\begin{thm}
A horizon is differentiable precisely at those points which belong to a single generator.
\end{thm}
We also note that horizons are null hypersurfaces whenever they are differentiable, so that the generators of a $C^2$ horizon are precisely the integral curves of the null vector fields:
\begin{prop}\label{TpH is a null hyperplane}
If a horizon $\mathscr H$ is differentiable at a point $p$, then $T_p\mathscr H$ is a null hyperplane.
\begin{proof}
Since $p$ belongs to a lightlike geodesic segment contained in $\mathscr H$, we know that $T_p\mathscr H$ contains null vectors. If $T_p\mathscr H$ were to contain a timelike vector, then there would be a timelike curve in $\mathscr H$ with this tangent vector.
This would contradict achronality of $\mathscr H$, and hence $T_p\mathscr H$ must be a null hyperplane.
\end{proof}
\end{prop}
Finally, we note that generators can only intersect in common endpoints.
\begin{prop}\label{Generators intersect only at endpoints}
Let $\mathscr H$ be a horizon, and suppose that $p$ is an interior point of a generator $\Gamma$. Then there is no other generator containing $p$.
\begin{proof}
Suppose that some other generator $\Gamma'$ contained $p$. Let $q$ be a point to the past of $p$ along $\Gamma'$, and let $r$ be a point to the future of $p$ along $\Gamma$. By following $\Gamma'$ from $q$ to $p$ and then $\Gamma$ from $p$ to $r$ we have connected $q$ and $r$ by a causal curve which not a null geodesic. 
By \cite[Proposition 2.6.9]{ChruscielElements} this curve cannot be achronal. Since the image of the curve belongs to $\mathscr H$, this contradicts achronality of $\mathscr H$.
\end{proof}
\end{prop}

\subsubsection{Cauchy horizons}
We now connect the statements in \autoref{Abstract horizons} about abstract horizons to the particular case of a Cauchy horizon in a spacetime. 
We begin by quoting \cite[Proposition 2.7]{GallowayNullGeometry}. A similar statement can be found in \cite[Proposition 2.10.6]{ChruscielElements}.
\begin{prop}
Let $S$ be an achronal subset of a spacetime $M$. Then the set $H^+(S) \setminus \edge(S)$, if nonempty, is an achronal $C^0$ hypersurface of $M$ ruled by null geodesics, each of which either is past inextendible in $M$ or has a past endpoint on $\edge(S)$.
\end{prop}

\begin{cor}\label{Cauchy horizons are horizons}
Let $M$ be a spacetime. Suppose that $S \subseteq M$ is an achronal set with $\edge(S) = \emptyset$. Then $H^+(S)$ is a horizon in the sense of \autoref{Definition of horizon}.
\begin{proof}
The proposition tells us that $H^+(S)$ is a topological hypersurface which is achronal and past null geodesically ruled.
To see that $H^+(S)$ is closed, note that it by definition is the difference of a closed set and an open set.
This completes the proof.
\end{proof}
\end{cor}

We conclude with a lemma allowing us to apply \autoref{Cauchy horizons are horizons} to closed spacelike hypersurfaces. The lemma follows from \cite[Lemma~8.3.3]{Kriele}.
\begin{lem}\label{Closed spacelike hypersurfaces are edgeless}
Let $M$ be a spacetime and let $S$ be a spacelike hypersurface which is closed as a set. Then $\edge(S) = \emptyset$.
\end{lem}

\subsubsection{Properties of nonsmooth horizons}
\begin{figure}
   \centering
   \includegraphics[width=0.7\textwidth]{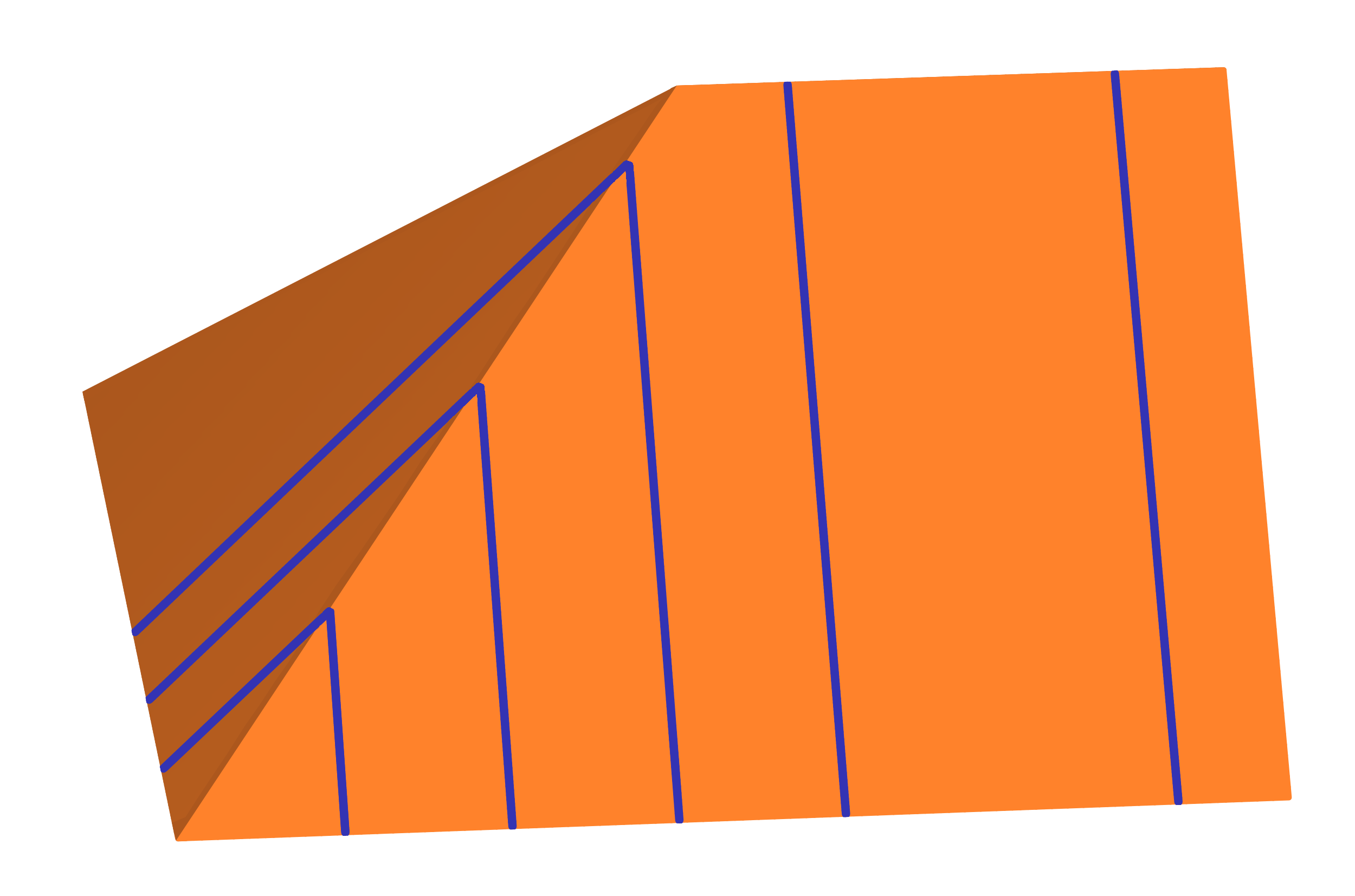}
   \caption{Part of the future Cauchy horizon of a spacelike rectangle in $(2+1)$-dimensional Minkowski space, with some of the generators shown. A more complicated example in which no open subset of the horizon is differentiable is given in \cite{ChruscielGalloway98}.}
   \label{fig:Non-smooth horizon}
\end{figure}
In general, Cauchy horizons are not $C^2$ hypersurfaces: \autoref{fig:Non-smooth horizon} shows an example of a non-$C^2$ Cauchy horizon. This particular example is ''almost $C^2$'' in the sense that it has a dense open subset which is $C^2$, so we would expect that many results about $C^2$ hypersurfaces are applicable to this example. However, it was shown in \cite{ChruscielGalloway98} that Cauchy horizons are not necessarily almost $C^2$. This means that the proofs of theorems like Tipler's theorem need to deal with horizons of lower regularity. For this reason, and in particular to prove \autoref{Compact horizons are smooth} about smoothness of compact Cauchy horizons, we need some results about general horizons. The definitions and results in this section can be found in \cite{CDGH01}.

\begin{defn}
Let $M$ be a spacetime, and let $(a, b) \times \Sigma \cong \mathcal O \subseteq M$ be an open subset such that each slice $\{t\} \times \Sigma$ is spacelike and each curve $(a, b) \times \{p\}$ is timelike. Let $N \subseteq M$ be a hypersurface. A function $f \colon \Sigma \to (a, b)$ is said to be a \emph{graphing function} of $N$ if $N \cap \mathcal O = \{(f(x), x) \mid x \in \Sigma\}$.
\end{defn}

Theorem~2.2 of \cite{CDGH01} says that any locally achronal hypersurface, in particular any horizon, is semi-convex (see \autoref{Semi-convexity}). This implies (see \cite[Proposition 2.1]{CDGH01}) that every point on the horizon has a globally hyperbolic spacetime neighborhood $(-a, a) \times \Sigma$ in which the horizon has a graphing function $f$ for which there is a subset $\Sigma\Al \subseteq \Sigma$ such that
\begin{itemize}
\item $\Sigma \setminus \Sigma\Al$ has measure zero,
\item $f$ is differentiable at all points of $\Sigma\Al$,
\item $f$ is \emph{twice-Alexandrov-differentiable} at all points $x \in \Sigma\Al$. In other words, there is a quadratic form $D^2f(x)$ such that for all $y \in \Sigma$
\[f(y)-f(x) - df(x)(y-x) = \frac{1}{2}D^2f(x)(x-y,x-y) + o(|x - y|^2).\]
\end{itemize}
Moreover, it is shown in \cite{CDGH01} that this notion is coordinate invariant: If $p = (f(x), x)$ with $x \in \Sigma\Al$ for one globally hyperbolic neighborhood of $p$, then $p = (\tilde f(\tilde x), \tilde x)$ with $\tilde x \in \tilde \Sigma\Al$ for any other neighborhood $(-\tilde a, \tilde a) \times \tilde \Sigma$ of $p$ with corresponding graphing function $\tilde f$ satisfying the above conditions. Hence the following definition makes sense.
\begin{defn}
Let $\mathscr H$ be a horizon in a spacetime. Denote by $\mathscr H\Al$ the set of all points $p \in \mathscr H$ which are images under a graphing function of one of the corresponding sets $\Sigma\Al$. We will call $\mathscr H\Al$ the set of \emph{Alexandrov points} of the horizon.
\end{defn}
\begin{remark}
By the definition of semi-convexity a semi-convex function is the sum of a $C^2$ function and a convex function, and hence locally Lipschitz. This means that horizons are Lipschitz hypersurfaces.
\end{remark}
Following \cite{CDGH01} we will now define the null mean curvature $\theta\Al$ and the null second fundamental form $B\Al$ of $\mathscr H\Al$. More precisely, we will define $\theta\Al$ and $B\Al$ on the intersection of $\mathscr H\Al$ with a globally hyperbolic coordinate neighborhood $\mathcal O$. This definition is \emph{not} coordinate invariant. However, $\theta\Al$ and $B\Al$ are defined up to pointwise scaling by a positive function, so the sign of $\theta\Al$ is globally well defined. 

\begin{defn}
Let $\mathscr H$ be a horizon in a spacetime $(M, g)$.
Choose a globally hyperbolic coordinate neighborhood $\mathcal O = (-a, a) \times \Sigma$ of some point in $\mathscr H$, and let $f \colon \Sigma \to (-a, a)$ be the graphing function of $\mathscr H$ in this neighborhood. Let the function $t \colon (-a, a) \times \Sigma \to (-a, a)$ be the projection. For each point $p \in \mathcal O \cap \mathscr H\Al$, with $x$ such that $p = (f(x), x)$, define $k(p) = -dt + df(x)$. This makes sense since $f$ is differentiable at all such points. Let $K$ be the vector field dual to $k$ with respect to $g$. Let $e_0 \in T_p\mathcal O$ be the vector which is $g$-dual to $dt$. Choose a basis $e_1, \ldots, e_n$ for $T_p\mathscr H$ such that
\begin{itemize}
\item $e_n = K_p$,
\item $g(e_i, e_i) = 1 \text{ if } 1 \leq i \leq n-1$,
\item $g(e_i, e_j) = 0 \text{ if } i \neq j$,
\item $g(e_i, e_0) = 0 \text{ if } 1 \leq i \leq n-1$.
\end{itemize}
We now define $\theta\Al$ and $B\Al$ using the coordinate formulae
\[\theta\Al = \sum_{l = 1}^{n-1} e_l^ie_l^j \left(D^2_{ij}f - \Gamma^\mu_{ij}k_\mu \right),\]
\[B\Al(X^ae_a, Y^be_b) = X^a Y^b e_a^i e_b^j \left(D^2_{ij}f - \Gamma^\mu_{ij}k_\mu \right).\]
In the definitions above, we have constructed an ''artificial'' covariant derivative $D^2_{ij} f - \Gamma^\mu_{ij}k_\mu$ using the Alexandrov second derivative $D^2f$ of $f$.
We emphasize again that this definition of $\theta\Al$ is not independent of the coordinate system. However, the definitions using different coordinate systems differ only by a positive multiplicative constant. In particular, the sign of $\theta\Al$ is invariantly defined (see \cite[Proposition 2.5]{CDGH01}).
\end{defn}
\begin{remark}
If $\mathscr H$ is $C^2$, then $\theta\Al$ and $B\Al$ agree with the null mean curvature $\theta_K$ and null second fundamental form $B_K$ defined in \autoref{The null Weingarten map}.
\end{remark}
\begin{remark}\label{Alexandrov Riccati equation}
By \cite[Theorem~5.1]{CDGH01} the $(1, 1)$-tensor $b\Al$ associated to $B\Al$ satisfies equation \eqref{Riccati equation for null Weingarten map} from \autoref{The null Weingarten map}.
\end{remark}
We will later derive a formula involving $\theta\Al$ for the area of a horizon, and knowing the sign of $\theta\Al$ will yield inequalities between different areas. In our case, the generators of the horizon will be past complete, so we will have use of the following result. It is a generalization of \cite[Proposition 4.17]{CDGH01}, and the proof of that proposition is sufficient for proving the generalization as well, since the proof considers one point at a time.
\begin{prop}\label{Sign of theta}
Let $M$ be a spacetime, and let $\mathscr H$ be a horizon in $M$. Let $A \subset \mathscr H\Al$ be a set of Alexandrov points of $\mathscr H$ such that each point in $A$ belongs to a generator which is complete to the past. Suppose that the null energy condition holds. Then
\[\theta\Al \leq 0 \quad \text{on $A$.}\]
\end{prop}
\begin{remark}
The result in \cite{CDGH01} is expressed with the opposite time orientation compared to our setting. Consequently we obtain the inequality $\theta\Al \leq 0$ instead of $\theta\Al \geq 0$.
\end{remark}

\subsection{A smoothness theorem}\label{A smoothness theorem}
In \cite[Section~4]{BudzynskiKondrackiKrolak01} the question was posed whether a compact Cauchy horizon is necessarily smooth. A negative answer was given by the same authors in \cite[Section~4]{BudzynskiKondrackiKrolak03}, where it was also mentioned that compactness together with some energy condition might be sufficient to guarantee smoothness. In this section, we show using methods from \cite{CDGH01} that this is indeed the case. When our proofs parallel those in \cite{CDGH01}, we will adhere to the notation in \cite{CDGH01}.

\subsubsection{Outline of the proof}
The theorem which will be proved in this section is \autoref{Compact horizons are smooth}, stating that compact Cauchy horizons in a spacetime which satisfies the null energy condition are smooth. 
We first give an outline of the proof. 
Horizons are Lipschitz null hypersurfaces, and so differentiable almost everywhere. At the points of differentiability there is a unique (up to scaling) null tangent vector, giving rise to an almost everywhere defined vector field on the horizon. By restricting to a suitably chosen subset of the horizon, we can define a flow along this vector field. One may then construct a $C^{1, 1}$ manifold containing, locally, this chosen subset, and extend the flow to a Lipschitz flow on the $C^{1,1}$ manifold. This is sufficient regularity to express how the area of a set changes under the flow, and this change of area is the central idea of the proof. To measure area, we introduce a Riemannian metric $\sigma$ on the spacetime. With a suitably chosen such metric, the change in area is related to the Alexandrov null mean curvature $\theta\Al$ of the Cauchy horizon. 
The argument for this relation between area change and $\theta\Al$ proceeds via a $C^2$ approximation of a part of the local $C^{1, 1}$ approximation of the original horizon.
Once the relation between $\theta\Al$ and area change has been established, knowledge of the sign of $\theta\Al$ gives an inequality for area change under the flow. A sufficient condition under which the sign of $\theta\Al$ may be determined is that all null geodesics in the horizon are complete in the past direction, together with an energy condition. \autoref{H+complete} tells us that the generators are complete. By these arguments, we determine that the flow increases area. However, the flow maps a subset of the horizon into itself, thereby decreasing area. Hence the only possibility is that the flow conserves area. We show in \autoref{No endpoints} that this implies that the horizon is smooth. 

\subsubsection{Flow sets and generator flow}
We wish to generalize the notion of the generator flow on $C^2$ null hypersurfaces discussed in \autoref{Generator flow on C2 null hypersurfaces} to possibly nonsmooth horizons. In other words, we want a flow along generators of a horizon $\mathscr H$. However, since some points belong to several generators it is in general not possible to do this on all of $\mathscr H$. Instead, we construct a smaller subset on which to define the flow.

\begin{defn}
Let $\mathscr H$ be a horizon in a spacetime $M$. 
Define the \emph{total flow set} of $\mathscr H$ to be the set $A_0(\mathscr H)$ of points $p \in \mathscr H$ such that the following conditions are satisfied:
\begin{itemize}
\item There is a unique generator $\Gamma$ of $\mathscr H$ passing through $p$.
\item The point $p$ belongs to the interior of $\Gamma$.
\item Each interior point of $\Gamma$ is an Alexandrov point.
\end{itemize}
Let $\sigma$ be a Riemannian metric on $M$. For $\delta > 0$ define the \emph{$\delta$-flow set} of $\mathscr H$ with respect to $\sigma$ to be the set
\[\begin{split}
A_\delta(\mathscr H, \sigma) = \{p \in A_0(\mathscr H) \mid	&\text{ The generator through $p$ exists}\\
								&\text{ for a $\sigma$-distance greater than $\delta$}\\
								&\text{ to the past and to the future}\}.
\end{split}\]
\end{defn}

\begin{remark}
Note that the total flow set is the union of all $\delta$-flow sets:
\[A_0(\mathscr H) = \bigcup_{\delta > 0} A_\delta(\mathscr H, \sigma).\]
\end{remark}
\begin{remark}
When the context allows it, we will sometimes drop $\mathscr H$ and $\sigma$ from the notation and write $A_0$ or $A_\delta$.
\end{remark}

For the next proposition, we will need the following result, a proof of which can be found in \cite[Theorem~5.6]{CDGH01}.
\begin{lem}\label{Almost all generators are Alexandrov}
Let $\mathscr H$ be a horizon in a spacetime $M$ of dimension $n+1$. Suppose that $\mathcal S$ is a $C^2$ hypersurface intersecting $\mathscr H$ properly transversally (in the sense that if $q \in \mathcal S \cap \mathscr H$ and the tangent space $T_q\mathscr H$ exists then $T_q\mathcal S$ is transverse to $T_q\mathscr H$). Define
\[S_0 = \{q \in \mathcal S \cap \mathscr H \mid \text{$q$ is an interior point of a generator of $\mathscr H$}\},\]
\[\begin{split}
S_1 = \{q \in S_0 \mid &\text{ all interior points of the generator}\\
&\text{ through $q$ are Alexandrov points of $\mathscr H$}\}.
\end{split}\]
Then $S_1$ has full $(n-1)$-dimensional Hausdorff measure in $S_0$.
\end{lem}

\begin{prop}\label{Union of flow sets has full measure}
Let $\mathscr H$ be a horizon in a spacetime $(M, g)$ of dimension $n+1$, and let $\sigma$ be a Riemannian metric on $M$. Let $\mathfrak h^n$ be the $n$-dimensional Hausdorff measure induced by the distance function induced by $\sigma$.
Then the total flow set $A_0$ of $\mathscr H$ has full $\mathfrak h^n$-measure in the sense that
\[\mathfrak h^n(\mathscr H \setminus A_0) = 0.\]
\begin{proof}
To show that $\mathscr H \setminus A_0$ has measure zero, it is sufficient to show that each point $p \in \mathscr H$ has an open neighborhood $U \subseteq M$ such that $\mathfrak h^n\left(U \cap (\mathscr H \setminus A_0)\right) = 0$, for $\mathscr H$ can be covered by countably many such neighborhoods since it is second-countable. The idea of the proof is to construct a spacetime of one dimension greater than $M$ and apply \autoref{Almost all generators are Alexandrov} in this higher-dimensional spacetime.

To this end, choose a globally hyperbolic neighborhood $U \subseteq M$ of $p \in \mathscr H$ diffeomorphic to $(-a, a) \times \Sigma$, where $\Sigma \subseteq \R^n$ and each slice $\{t\} \times \Sigma$ is spacelike. We may choose the zero slice to be such that $p \in \{0\} \times \Sigma$. Let $\widehat M = M \times I$ denote the product manifold which is equipped with the metric $\widehat g = g + ds^2$, where $s$ refers to the coordinate in the open interval $I$. Let $\widehat{\mathscr H} = \mathscr H \times I$. Let $\pi \colon \widehat M \to M$ denote the projection. We now verify that $\widehat{\mathscr H}$ is a horizon in $\widehat M$. Since $\mathscr H$ is an embedded topological hypersurface, so is $\widehat{\mathscr H}$. By definition of the product topology, $\widehat{\mathscr H} = \pi^{-1}(\mathscr H)$ is closed. If there were some timelike curve $\gamma$ between two points of $\widehat{\mathscr H}$, then $\pi \circ \gamma$ would be a timelike curve between two points of $\mathscr H$ contradicting achronality of $\mathscr H$, so $\widehat{\mathscr H}$ must also be achronal. To see that $\pi \circ \gamma$ is indeed timelike, note that by definition of $\widehat g$ it holds that $g(\pi_* V, \pi_* V) \leq \widehat g(V, V)$ for all vectors $V$. Finally $\widehat{\mathscr H}$ is past null geodesically ruled since if $\Gamma$ is a past inextendible null $M$-geodesic contained in $\mathscr H$, then $\Gamma \times \{s\}$ is a past inextendible null $\widehat M$-geodesic contained in $\widehat{\mathscr H}$ for each $s \in I$.

We now wish to construct, after possibly decreasing $a$ or shrinking $I$, a diffeomorphism $\rho \colon I \to (-a, a)$ such that the hypersurface
\[\mathcal S := \left\{(t, q, s) \in (-a, a) \times \Sigma \times I \mid t = \rho(s)\right\}\]
is spacelike. A possible choice of basis for the tangent space $T_{(t,q,s)}\mathcal S \subseteq T_q\R \times \Sigma \times \R$ of $\mathcal S$ at some point $(t, q, s)$ consists of a basis for the tangent space of $\Sigma$ together with the vector $(\rho'(s), 0, 1)$. The basis of $T_q\Sigma$ consists of spacelike vectors since $\Sigma$ is spacelike, and if $\rho'(s)$ is sufficiently close to zero then $(\rho'(s), 0, 1)$ is also spacelike (since $(0, 0, 1)$ is spacelike by definition of $\widehat g$, and the set of spacelike vectors at a point is open). This means that for each $(t, q) \in (-a, a) \times \Sigma$, there is some $c(t, q) > 0$ such that if $\zeta < c(t, q)$ then for any $s \in I$ the vector $(0, \zeta, 1) \in T_{(t, q, s)}$ is spacelike. Since $g$ is smooth, $c$ can be chosen smooth. Hence $c$ takes some minimum on every compact subset of $(-a, a) \times \Sigma$, This minimum is positive since $c$ is positive on $(-a, a) \times \Sigma$, after possibly shrinking $\Sigma$ and $a$. We may then find some real number $\zeta$ such that $0 < \zeta < c(t, q)$ for all $(t, q) \in (-a, a) \times \Sigma$. Letting $\rho(s) = \zeta(s - s_0)$ where $s_0$ is the midpoint of $I$, and subsequently shrinking $I$ or $a$ to make $\rho$ bijective, we have found a diffeomorphism $\rho$ making $\mathcal S$ spacelike.
Since $\rho$ is a diffeomorphism, the restriction of the projection $\pi \colon \widehat M \to M$ to $\mathcal S$ is also a diffeomorphism.

Now $\mathcal S$ is a smooth hypersurface in $\widehat M$, which intersects $\widehat{\mathscr H}$ properly transversally in the sense that if $q \in \mathcal S \cap \widehat{\mathscr H}$ and the tangent space $T_q\widehat{\mathscr H}$ exists then $T_q\mathcal S$ is transverse to $T_q\widehat{\mathscr H}$. Let
\[\widehat{S_0} = \{q \in \mathcal S \cap \widehat{\mathscr H} \mid q \text{ is an interior point of a generator of } \widehat{\mathscr H}\},\]
\[S_0 = \{q \in U \cap \mathscr H \mid q \text{ is an interior point of a generator of } \mathscr H\}.\]
Note that $\pi(\widehat{S_0}) = S_0$ since if $p$ is an interior point of a generator $\Gamma$ then $\pi(p)$ is an interior point of the generator $\pi(\Gamma)$ and vice versa. Note further that it holds that ${\pi(\mathcal S \cap \widehat{\mathscr H}) = U \cap \mathscr H}$.
Moreover, the projection $\pi$ restricted to $\mathcal S$ is bijective and hence $\pi((\mathcal S \cap \widehat{\mathscr H}) \setminus \widehat{S_0}) = (U \cap \mathscr H) \setminus S_0$. 

Since $\pi$ restricted to $\mathcal S$ is a diffeomorphism, both $\left.\pi\right|_{\mathcal S}$ and its inverse $\left(\left.\pi\right|_{\mathcal S}\right)^{-1}$ are locally Lipschitz so that $\mathfrak h^n((\mathcal S \cap \widehat{\mathscr H}) \setminus \widehat{S_0}) = 0$ if and only if $\mathfrak h^n((U \cap \mathscr H) \setminus S_0) = 0$. The latter set $(U \cap \mathscr H) \setminus S_0$ is the set of endpoints of generators of $\mathscr H$ contained in $U$. It is shown in \cite[Theorem~3.5]{BeemKrolak98} and \cite[Theorem~1]{CFGH02} that this set has zero $\mathfrak h^n$-measure. This means that we can conclude that $\mathfrak h^n\left((\mathcal S \cap \widehat{\mathscr H}) \setminus \widehat{S_0}\right) = 0$. In other words, $\widehat{S_0}$ has full measure in $\mathcal S \cap \widehat{\mathscr H}$.

Let
\[\begin{split}
\widehat{S_1} = \{q \in \widehat{S_0} \mid &\text{ all interior points of the generator}\\
				&\text{ through $q$ are Alexandrov points of $\widehat{\mathscr H}$}\}.
\end{split}\]
By \autoref{Almost all generators are Alexandrov} the set $\widehat{S_1}$ has full $\mathfrak h^n$ measure in $\widehat{S_0}$. Hence it also has full $\mathfrak h^n$-measure in $\mathcal S \cap \widehat{\mathscr H}$. Since $\pi$ is bi-Lipschitz, $\pi(\widehat{S_1})$ has full $\mathfrak h^n$-measure in $U \cap \mathscr H$.

The projection $\pi \colon \widehat M \to M$ maps generators to generators, and Alexandrov points of $\widehat{\mathscr H}$ to Alexandrov points of $\mathscr H$, so each point of $\widehat{S_1}$ belongs to $A_0$. We have then shown that $A_0 \cap U$ contains a subset $\pi(\widehat{S_1})$ which has full measure in $\mathscr H \cap U$. Hence $A_0$ itself has full measure in $\mathscr H \cap U$. As noted in the beginning of the proof, $\mathscr H$ may be covered by countably many such sets $U$, so we have shown that $A_0$ has full $\mathfrak h^n$-measure in $\mathscr H$. This completes the proof.
\end{proof}
\end{prop}

\begin{defn}
Let $\mathscr H$ be a horizon in a spacetime $(M, g)$, and let $A_0$ be its total flow set. Let $\sigma$ be a Riemannian metric on $M$. Since $\mathscr H$ is differentiable at all points in $A_0$, there is a unique $\sigma$-unit past-directed null vector tangent to $\mathscr H$ at each point in $A_0$. This defines a vector field $T$ on $A_0$, which is tangent to the generators of $\mathscr H$. Recall that $A_0$ contains full generators, and hence full integral curves of $T$. We will call the flow of $T$ the \emph{generator flow} of $\mathscr H$ with respect to $\sigma$, and denote it by $(t, p) \mapsto \beta_t(p)$. 
\end{defn}
Note that $\beta_t$ is not in general defined on all of $A_0$ for any $t > 0$. However, it will be defined on all of $A_0$ for all $t > 0$ in the case considered in our main theorem, so we will mainly be concerned with this case.

\subsubsection{Generator flow is area-preserving}
The purpose of this section is to prove that the generator flow on a horizon with respect to a certain family of Riemannian metrics preserves the associated Hausdorff measure if the null mean curvature is nonpositive. Our first goal is to construct a $C^{1, 1}$ approximation of the horizon to be able to express the volume change. We do this in \autoref{N construction} and \autoref{Lipschitz extension}. We then construct a $C^2$ approximation of the horizon to compute the volume change in \autoref{Local volume change under flow} and \autoref{Area unchanged by flow}. The complicated constructions necessary are contained in \autoref{Inequality for J}.

We begin by stating an extension result, which is proved in \cite[Proposition 6.6]{CDGH01}.
\begin{lem}\label{Whitney}
Let $B \subseteq \R^n$ be an arbitrary subset and $f \colon B \to \R$ be an arbitrary function. Suppose that there is some constant $C > 0$, and some function $B \to \R^n$, $p \mapsto a_p$, (not necessarily continuous) such that the following two conditions hold:
\begin{enumerate}
\item $f$ has global upper and lower support paraboloids of opening $C$. Explicitly, for all $x, p \in B$,
\[\left| f(x) - f(p) - \langle x - p, a_p \rangle \right| \leq C||x - p||^2.\]
\item The upper and lower support paraboloids of $f$ are disjoint. Explicitly, for all $p,q \in B$ and all $x \in \R^n$,
\[f(p) + \langle x - p, a_p \rangle - C||x - p||^2 \leq f(q) + \langle x - q, a_q \rangle + C||x - q||^2.\]
\end{enumerate}
 Then there is a function $F \colon \R^n \to \R$ of class $C^{1, 1}_{loc}$ such that $f$ is the restriction of $F$ to $B$.
\end{lem}

Using this lemma, we may prove the following.

\begin{lem}\label{N construction}
Let $\mathscr H$ be a horizon in an $(n+1)$-dimensional spacetime $M$. Let $\sigma$ be any Riemannian metric on $M$, let $\delta > 0$ and let $A_\delta$ be the $\delta$-flow set of $\mathscr H$ with respect to $\sigma$.
Let $p \in A_\delta$. Then there is some open globally hyperbolic neighborhood $V \subseteq M$ of $p$ and a $C^{1, 1}$ hypersurface $N \subseteq V$ in $M$ such that $A_\delta \cap V \subseteq N$.
\begin{proof}
For each point $q \in A_\delta$, let $q^+$ denote the point a $\sigma$-distance $\delta$ to the future along the unique generator through $q$. Similarly, let $q^-$ denote the point along the generator a distance $\delta$ to the past. By one of the defining properties of $A_\delta$, we have $q^+, q^- \in \mathscr H$.

By the same reasoning as is used in the proof of Lemma~6.9 in \cite{CDGH01}, one may obtain a globally hyperbolic neighborhood $V \subseteq W$ of $p$ and a constant $C > 0$ with the following properties:
\begin{itemize}
\item $V$ is diffeomorphic to $(-a, a) \times B^n(r)$ with the slices $\{t\} \times B^n(r)$ spacelike and the curves $(-a, a) \times \{x\}$ timelike and future-directed for all $t \in (-a, a)$ and all $x \in B^n(r)$.
\item Let $f$ denote the graphing function of the horizon over $B^n(r)$, i.e. the function such that $V \cap \mathscr H = \{(f(x), x) \mid q \in B^n(r)\}$. For each $q = (f(x_q), x_q) \in V \cap A_\delta$, the graph of the function
\[f^-_q(x) = f(x_q) + df(x_q)(x - x_q) - C||x - x_q||^2,\]
with the exception of the point $q = (f(x_q), x_q)$ itself, lies in the timelike past $I^-(q^+, V)$ of $q^+$.
\item For each $q = (f(x_q), x_q) \in V \cap A_\delta$, the graph of the function
\[f^+_q(x) = f(x_q) + df(x_q)(x - x_q) + C||x - x_q||^2,\]
with the exception of the point $q = (f(x_q), x_q)$ itself, lies in the timelike future $I^+(q^-, V)$ of $q^-$.
\end{itemize}
Note that if this holds for some value of $C$, it holds for all larger values of $C$ as well.

We will now show that these conditions imply the first hypothesis of \autoref{Whitney}. Suppose that the condition is violated. Then either $f(x) > f_q^+(x)$ or $f(x) < f_q^-(x)$ for some $q = (f(x_q), x_q) \in A_\delta$ and $x \in B^n(r)$. The argument is the same for both cases, so suppose without loss of generality that the first is the case. Since $f(x_q) = f_q^+(x_q)$ we must have $x \neq x_q$. Then $(f_q^+(x), x)$ belongs to the timelike future of $q^-$, by the choice of $C$.
However, since $f(x) > f_q^+(x)$, the point $(f(x), x)$ lies to the timelike future of $(f_q^+(x), x)$. This means that we can connect $q^-$ to $(f_q^+(x), x)$ to $(f(x), x)$ by a timelike curve. Hence $(f(x), x)$ belongs to the timelike future of $q^-$. Since both points belong to the horizon, this violates achronality of the horizon. This proves the first hypothesis of \autoref{Whitney}.

For the second hypothesis, note that the first continues to hold if we increase $C$. By making sure that $C$ is sufficiently large compared to the Lipschitz constant of $f$ and the values of $f$, one may conclude as in the proof of Lemma~6.9 in \cite{CDGH01} that the second hypothesis is satisfied as well.

Let $B$ denote the projection of $A_\delta \cap V$ on $B^n(r)$. We can then apply the extension theorem described in \autoref{Whitney} to obtain a $C^{1,1}$ extension $\R^n \to \R$ of $\left. f \right|_B \colon B \to (-a, a)$. Let $F \colon \R^n \to \R$ denote the restriction to $B^n(r)$ of this extension. By definition $F$ agrees with $f$ on $B$. In particular, the graph of $F$ contains $p = (f(x_p), x_p)$, so $F(x_p) \in (-a, a)$. Since $F$ is continuous, there is some neighborhood $B^n(\epsilon) \subseteq B^n(r)$ of $x_p$ such that $F(B^n(\epsilon)) \subseteq (-a, a)$. Hence by shrinking the neighborhood $V$ to $(-a, a) \times B^n(\epsilon)$ and letting $N$ be the graph of $F$ there, we have obtained a $C^{1, 1}$ hypersurface containing $A_\delta \cap V$.
\end{proof}
\end{lem}

\begin{lem}\label{Lipschitz extension}
Let $\mathscr H$ be a horizon in an $(n+1)$-dimensional spacetime $(M, g)$ equipped with a Riemannian metric $\sigma$, let $\delta > 0$, let $A_\delta$ be the $\delta$-flow set of $\mathscr H$ with respect to $\sigma$, let $\tilde A_\delta$ be the full-density subset (in the sense of \autoref{Riemannian full-density subset}) of $A_\delta$, let $V$ be a globally hyperbolic open neighborhood of $p$ and let $N \subseteq V$ be a $C^{1, 1}$ hypersurface, containing $A_\delta \cap V$, which can be represented by a graphing function in $V$.

Fix $t \geq 0$. Let $\beta_t \colon A_\delta \cap V \to A_0$ be the restriction of the generator flow (with respect to $\sigma$) to $A_\delta \cap V$, and suppose that this flow is defined on all of $A_\delta \cap V$. Then there is a neighborhood $U \subseteq V$ of $p$ such that the restriction of $\beta_t$ to $\tilde A_\delta \cap U$ is the restriction of a locally Lipschitz function $\widehat{\beta_t} \colon N \cap U \to M$.
\begin{proof}
In the trivial case $t = 0$ we can let $\widehat \beta_t$ be the identity on $N$. Hence we can assume for the remainder of the proof that $t > 0$.

Let $(a, b) \times \Sigma$ be a decomposition in space and time of the globally hyperbolic neighborhood $V$ of $p$, and let $f$ denote the graphing function of $N$ with respect to this decomposition.

We wish to construct a Lipschitz vector field normal to $N$ in a neighborhood of $p = (f(x), x)$. Choose a frame $(e_i)_{i = 1}^n$ close to $p$ consisting of the pushforward of a frame of $\Sigma$ close to $x$ under the map $y \mapsto (f(y), y)$. This frame is Lipschitz since $f$ is $C^{1,1}$. By shrinking $V$ we may assume that the frame covers all of $N$. The condition that a vector field $\mathbf n$ along $N$ is normal to $N$ with respect to the spacetime metric $g$, and consists of unit vectors with respect to the Riemannian metric $\sigma$ can be expressed by saying that $\mathbf n$ satisfies the $n+1$ equations
\[\begin{split}
\sigma(\mathbf n, \mathbf n) - 1 &= 0,\\
g(e_1, \mathbf n) &= 0\\
g(e_2, \mathbf n) &= 0\\
\vdots&\\
g(e_n, \mathbf n) &= 0.
\end{split}\]
Choose a trivialization $N \times \R^{n+1}$ of $\left.TM\right|_N$. Define $F \colon N \times \R^{n+1} \to \R^{n+1}$ by the above equations. Explicitly
\[F(\mathbf n) = (\sigma(\mathbf n, \mathbf n) - 1, g(e_1, \mathbf n), \ldots, g(e_n, \mathbf n)).\]
Let $\mathbf n$ be a zero of $F$. The tangent map of $F$ at $\mathbf n$ with respect to the $\R^{n+1}$ component is
\[\mathbf k \mapsto (2\sigma(\mathbf n, \mathbf k), g(e_1, \mathbf k), \ldots, g(e_n, \mathbf k)).\]
We wish to show that this tangent map has full rank. For dimensional reasons, this is equivalent to its kernel being trivial. If $\mathbf k$ belongs to the kernel, then $g(e_i, \mathbf k) = 0$ for all $1 \leq i \leq n$. Hence $\mathbf k$ is a normal vector to $N$, and hence parallel to $\mathbf n$. If $\mathbf k$ belongs to the kernel of the tangent map then it also holds that $\sigma(\mathbf n, \mathbf k) = 0$. When $\mathbf k$ is parallel to $\mathbf n$, this can only happen when $\mathbf k = 0$. 
This shows that the tangent map of the $\R^{n+1}$ component of $F$ has full rank at zeros of $F$.
Clearly there is a vector at $p$ which is a zero of $F$; simply take a normal vector and rescale it. This means that we (after choosing a local trivialization of $\left.TM\right|_N$ around $p$) can apply Clarke's Implicit Function Theorem (Corollary, p.\,256 in \cite{Clarke}) to conclude that there is a Lipschitz function $\mathbf n$ satisfying $F(q, \mathbf n(q)) = 0$ in a neighborhood of $p$. By shrinking $V$ if necessary, we have then found a Lipschitz normal (with respect to $g$) vector field to $N$ which is of unit length (with respect to $\sigma$). By shrinking $V$ further and replacing $\mathbf n$ with $-\mathbf n$ if necessary, we may assume that $\mathbf n$ is past-directed wherever it is causal.

By considering graphing functions of $N$ and $\mathscr H$ and applying the result about tangent spaces at full-density points described in \autoref{Density differentials} we see that the tangent spaces $T_qN$ and $T_q\mathscr H$ agree at all $q \in \tilde A_\delta$.
Consider now a point $q \in \tilde A_\delta$.
By \autoref{TpH is a null hyperplane} the tangent space $T_q\mathscr H$ is a null hyperplane. Since $q \in \tilde A_\delta$ we have $T_qN = T_q \mathscr H$, so $T_qN$ is also a null hyperplane. The normal vector $\mathbf n_q$ to the null hyperplane $T_qN$ is then null, and any two null vectors in a null hyperplane are parallel, so the vector $\mathbf n_q$ for a point $q \in \tilde A_\delta$ is parallel to any tangent vector of the null geodesic generator through $q$.

Once again, consider some point $q \in \tilde A_\delta \subseteq N$. Since $\beta_t(q)$ is a point along the geodesic (with respect to the spacetime metric $g$) with initial velocity parallel to $\mathbf n_q$ (which is nonzero and past-directed) it holds that there is some function $r \colon \tilde A_\delta \to (0, \infty)$ such that
\[\beta_t(q) = \exp^g(r(q) \mathbf n_q).\]

For each $q \in N$ there is a unique positive real number $\hat r(q)$ such that the $\sigma$-distance between $q$ and $\exp^g(\hat r(q) \mathbf n_q)$ along the curve $\tau \to \exp^g(\tau \mathbf n_q)$ is precisely $t$. By definition $r$ and $\hat r$ coincide on $\tilde A_\delta$.

We now want to use Clarke's implicit function theorem (Corollary, p.\,256 in \cite{Clarke}) again, this time to conclude that $\hat r$ is locally Lipschitz. By definition, the choice $\xi = \hat r$ solves the equation
\[\forall q \in N \quad \int_0^1 \left|\left|\partderiv{}{\tau} \exp^g(\tau \xi(q) \mathbf n_q)\right|\right|_\sigma \ d\tau = 1,\]
and this solution is of course unique if we require that $\xi(q) > 0$ everywhere.
In other words $\xi = \hat r$ is the unique positive function satisfying $F(q, \xi(q)) = 0$ where $F \colon N \times \R \to \R$ is defined by
\[F(q, t) = \left(\int_0^1 \left|\left|\partderiv{}{\tau} \exp^g(\tau t \mathbf n_q)\right|\right|_\sigma \ d\tau\right) - 1.\]
Since $\mathbf n$ is locally Lipschitz, so is $F$.
Note that $F$ has a partial derivative with respect to $t$ and that $\partderiv{F}{t} \neq 0$ everywhere since $\mathbf n$ is nowhere zero.
Clarke's implicit function theorem now tells us that there is a solution $\xi$ of $F(q, \xi(q)) = 0$ with $\xi(p) = r(p)$ which is Lipschitz in a neighborhood of $p$. Since $r(p)$ is positive, so is $\xi$ in a neighborhood of $p$. Since we already know that the only positive solution of this equation is $\hat r$, this shows that $\hat r$ is Lipschitz in some neighborhood $U$ of $p$.

Since $\hat r$ is Lipschitz on $U$, the function $\widehat{\beta_t} \colon N \cap U \to M$ defined by
\[\widehat{\beta_t}(q) = \exp^g(\hat r(q) \mathbf n_q)\]
is also Lipschitz. The restriction of this function to $\tilde A_\delta \cap U$ agrees with $\beta_t$, completing the proof.
\end{proof}
\end{lem}

For future reference, we note the following corollary.
\begin{cor}\label{Flowed A_delta has full measure}
Let $\tilde A_\delta$ be the full-density subset of a $\delta$-flow set $A_\delta$, and let $U$ be any set such that the generator flow with respect to some Riemannian metric $\sigma$ is defined on all of $U \cap A_\delta$. Let $\mathfrak h^n$ be the $n$-dimensional Hausdorff measure associated to $\sigma$. Then $\beta_t(\tilde A_\delta \cap U)$ has full $\mathfrak h^n$-measure in $\beta_t(A_\delta \cap U)$.
\begin{proof}
When $U$ is contained in a sufficiently small open set, \autoref{Lipschitz extension} tells us that $\beta_t$ is the restriction of a Lipschitz function. Hence $\beta_t(A_\delta \cap U) \setminus \beta_t(\tilde A_\delta \cap U)$ has $\mathfrak h^n$-measure zero, since $A_\delta \setminus \tilde A_\delta$ has $\mathfrak h^n$-measure zero. If $U$ is not sufficiently small, it may be covered by countably many such small open sets since it is second-countable, giving the same conclusion.
\end{proof}
\end{cor}

\begin{prop}\label{Local volume change under flow}
Let $\mathscr H$ be a horizon in an $(n+1)$-dimensional spacetime $(M, g)$ equipped with a Riemannian metric $\sigma$ of the form
\[\sigma(X, Y) = g(X, Y) + 2g(X, V)g(Y, V)\]
for some timelike $g$-unit vector field $V$.
Let $\mathfrak h^n$ be the $n$-dimensional Hausdorff measure associated to $\sigma$, let $\delta > 0$, let $A_\delta$ be the $\delta$-flow set, let $\tilde A_\delta$ be the full-density subset of $A_\delta$, let $t > 0$, let $\beta_t$ be the restriction to $A_\delta$ of the generator flow with respect to $\sigma$. Suppose that $\beta_t$ is defined on all of $A_\delta$ and that $\theta\Al \leq 0$ on all of $\mathscr H\Al$.

Then every $p \in A_\delta$ has a neighborhood $Z$ which is open in $A_\delta$ such that there is a measurable function $\Psi$ with $\Psi \geq 1$ almost everywhere such that
\[\int_{\tilde A_\delta} \varphi \Psi d\mathfrak h^n = \int_{\beta_t(\tilde A_\delta)} \varphi(\beta_t^{-1}(y)) d\mathfrak h^n(y).\]
for every $\varphi \colon A_\delta \to \R$ which is $\mathfrak h^n$-integrable and supported in $Z \cap \tilde A_\delta$.
Moreover, if $\Psi = 1$ almost everywhere on $Z$, then $\theta\Al = 0$ almost everywhere on $Z$.

\begin{remark}
A sufficient condition for having $\theta\Al \leq 0$ is that the generators of $\mathscr H$ are complete in the past direction, together with the null energy condition (see \autoref{Sign of theta}).
\end{remark}
\begin{proof}
Choose some point $p \in A_\delta$. By \autoref{N construction}, there is a globally hyperbolic open spacetime neighborhood $U$ of $p$ and a $C^{1, 1}$ hypersurface $N \subseteq U$ such that $A_\delta \cap U \subseteq N$. \autoref{Lipschitz extension} tells us that after possibly shrinking $U$, the generator flow $\beta_t \colon A_\delta \to A_0$ is the restriction of a Lipschitz function $\widehat{\beta_t} \colon N \cap U \to M$.

Let $Z = A_\delta \cap U$.
Let $\varphi \colon A_\delta \to \R$ be $\mathfrak h^n$-integrable and supported in $Z \cap \tilde A_\delta$. Since $\widehat{\beta_t}$ is Lipschitz on $N$, Theorem~3.1 of \cite{Federer59} tells us that
\[\int_N \varphi J(\widehat{\beta_t}) d\mathfrak h^n = \int_N \left(\sum_{x \in \widehat{\beta_t}^{-1}(y)} \varphi(x)\right) d\mathfrak h^n(y).\]
Here $J(\widehat{\beta_t})$ is the Jacobian determinant of $\widehat{\beta_t}$ with respect to $\sigma$ at points where $\widehat{\beta_t}$ is differentiable. Since $\widehat{\beta_t}$ is Lipschitz, $J(\widehat{\beta_t})$ is thus almost everywhere defined on $N$, which is sufficient for the integral to make sense.

Note that $\varphi$ is zero outside of $Z \cap \tilde A_\delta$, so
\[\sum_{x \in \widehat{\beta_t}^{-1}(y)} \varphi(x) = \sum_{x \in \widehat{\beta_t}^{-1}(y) \cap Z \cap \tilde A_\delta} \varphi(x).\]
Note that $\widehat{\beta_t}^{-1}(y) \cap Z \cap \tilde A_\delta$ is the inverse image of $y$ under the restriction of $\widehat{\beta_t}$ to $Z \cap \tilde A_\delta$. This restriction agrees with the restriction of $\beta_t$ to the same set. Since $\beta_t$ is injective on $A_\delta$, its restriction to $Z \cap \tilde A_\delta$ is injective as well. 
This means that
\[\widehat{\beta_t}^{-1}(y) \cap Z \cap \tilde A_\delta = \begin{cases}
\{\beta_t^{-1}(y)\} \quad & \text{if } y \in \beta_t(Z \cap \tilde A_\delta),\\
\emptyset \quad & \text{otherwise.}
\end{cases}\]
Using this together with the fact that $\varphi$ is zero outside of $Z$ we see that
\[\begin{split}
\sum_{x \in \widehat{\beta_t}^{-1}(y)} \varphi(x)
&=
\begin{cases}
\varphi(\beta_t^{-1}(y)) \quad & \text{if } y \in \beta_t(Z \cap \tilde A_\delta),\\
0 \quad & \text{otherwise,}
\end{cases}\\
&=
\begin{cases}
\varphi(\beta_t^{-1}(y)) \quad & \text{if } y \in \beta_t(\tilde A_\delta),\\
0 \quad & \text{otherwise.}
\end{cases}
\end{split}\]
Hence
\[\int_{\tilde A_\delta} \varphi J(\widehat{\beta_t}) d\mathfrak h^n = \int_{\beta_t(\tilde A_\delta)} \varphi(\beta_t^{-1}(y))d\mathfrak h^n(y).\]

To complete the proof of the theorem, we need to show that $J(\widehat{\beta_t}) \geq 1$ almost everywhere on $\tilde A_\delta \cap U$, and that $J(\widehat{\beta_t}) = 1$ almost everywhere on $\tilde A_\delta \cap U$ only if $\theta\Al = 0$ almost everywhere on $A_\delta \cap U$, possibly after shrinking $U$. We can then choose $\Psi$ to be $J(\widehat{\beta_t})$. Since the argument for proving these statements is quite long, we prove them separately as \autoref{Inequality for J}.
\end{proof}
\end{prop}

\begin{lem}\label{Inequality for J}
Fix $t$. Let $p$, $U$, $A_\delta$, $\tilde A_\delta$, $N$ and $\widehat \beta_t$ be as in \autoref{Local volume change under flow}. After possibly shrinking $U$ to a smaller neighborhood of $p$, it holds that $J(\widehat{\beta_t}) \geq 1$ almost everywhere (with respect to $\mathfrak h^n$) on $A_\delta \cap U$. Moreover, if $C \subset U$ is a closed set in the subspace topology on $U$ and $J(\widehat{\beta_t}) = 1$ almost everywhere on $A_\delta \cap C$ then $\theta\Al = 0$ almost everywhere on $A_\delta \cap C$.
\begin{proof}
\begin{proofpart}{Construction of a set $\widehat B$}
After possibly shrinking $U$ to a smaller neighborhood of $p$ and decomposing it as $U = (a, b) \times \Sigma$ (with the curves $(a, b) \times \{q\}$ timelike and the slices $\{\tau\} \times \Sigma$ spacelike) where $\Sigma$ is an open subset of $\R^n$, we may view $\mathscr H$ as the graph of a semi-convex function $f$, as noted in \autoref{Abstract horizons}. Similarly, $N$ is the graph of a $C^{1,1}$ function $g$. Let $\mathscr L^n$ denote Lebesgue measure on $\Sigma$. By \cite[Theorem~3.1.15]{Federer69}, for each positive integer $k$ there is a $C^2$ function $g_k \colon \Sigma \to \R$ such that
\[\mathscr L^n\left(\{x \in \Sigma \mid g_k(x) \neq g(x)\}\right) < 1/k.\]

Let $\operatorname{pr} \tilde A_\delta$ be the projection of $\tilde A_\delta$ on $\Sigma$. Let $B$ be the full-density subset of $\operatorname{pr} \tilde A_\delta$. By \autoref{Full-density sets in Rn have full measure} the set $B$ has full $\mathscr L^n$-measure in $\operatorname{pr} \tilde A_\delta$.
Letting
\[B_k = B \cap \{x \in \Sigma \mid g_k(x) = g(x)\}\]
we then have
\[\mathscr L^n(B \setminus B_k) < 1/k.\]
Once again, we discard low-density points: Let $\tilde B_k$ be the full-density subset of $B_k$. 
Then $\tilde B_k$ has full measure in $B_k$ by \autoref{Full-density sets in Rn have full measure} so
\[\mathscr L^n(B \setminus \tilde B_k) < 1/k.\]
Let $\Sigma_{Rad}$ denote the points of $\Sigma$ where $g$ is twice differentiable in the sense that it has second order expansions of the form
\begin{equation}\label{Rademacher expansion}
\begin{split}
g(x) &= g(x_0) + dg(x_0)(x - x_0) + \frac{1}{2}D^2g(x_0)(x - x_0, x - x_0) + o(|x - x_0|^2),\\
dg(x) &= dg(x_0) + D^2g(x_0)(x - x_0, \cdot) + o(|x - x_0|).
\end{split}
\end{equation}
Rademacher's theorem tells us that since $g$ is $C^{1, 1}$, the set $\Sigma_{Rad}$ has full measure in $\Sigma$.
Defining
\[\widehat B_k := \tilde B_k \cap \Sigma_{Rad},\]
\[\widehat B := \bigcup_{k \in \N} \widehat B_k = \Sigma_{Rad} \cap \bigcup_{k \in \N} \tilde B_k\]
we then know that $\widehat B$ has full $\mathscr L^n$-measure in $\operatorname{pr} \tilde A_\delta$. Since $g$ is Lipschitz, the graph of $g$ over $\widehat B$ has full $\mathfrak h^n$-measure in $\tilde A_\delta \cap U$ by \autoref{Zero sets under Lipschitz maps}. It is now sufficient to show that $J(\widehat{\beta_t})(p_0) \geq 1$ whenever $p_0 = (g(x_0), x_0)$ is such that $x_0 \in \widehat B$.

Choose some $x_0 \in \widehat B$. Since $\widehat B \subseteq \operatorname{pr} \tilde A_\delta$, the functions $f$ and $g$ agree at $x_0$, and $x_0$ is an Alexandrov point of $f$. Hence we have the expansion
\[f(x) = f(x_0) + df(x_0)(x - x_0) + \frac{1}{2}D^2f(x_0)(x - x_0, x - x_0) + o(|x - x_0|^2).\]
Since $x_0 \in \widehat B$, we know that $g$ is twice differentiable at $x_0$ so that we have the expansions of equation \eqref{Rademacher expansion}.
Moreover, $x_0 \in B_i$ for some $i \in \N$ so $g_i(x_0) = g(x_0)$. Fix this value of $i$ for the remainder of the proof. Moreover, by definition of $\widehat B$, the point $x_0$ is a full-density point of $\operatorname{pr} A_\delta$, and $g$ and $f$ agree on $\operatorname{pr} A_\delta$. Similarly, $x_0$ is a full-density point of $B_i$, and $g$ and $g_i$ agree on $B_i$. This means that we can use \autoref{Density differentials} to conclude that
\begin{equation}\label{Common differentials}
df(x_0) = dg(x_0) = dg_i(x_0)
\end{equation}
and
\[D^2g_i(x_0) = D^2g(x_0) = D^2f(x_0).\]
\end{proofpart}

\begin{proofpart}{Construction of a $C^2$ null hypersurface $\mathscr H_i$}
Let $N_i$ denote the graph of $g_i$ over $\Sigma$. Let $\widehat N_i$ denote a $C^2$ spacelike hypersurface in $N_i$ containing $p_0$. Thus $\widehat N_i$ has codimension $2$ in the spacetime $M$. 
Let $\mathbf n_i$ denote the past-directed $\sigma$-unit normal null vector field of $\widehat N_i$ such that $\mathbf n_i(p_0)$ is the $\sigma$-unit tangent of the (unique since $p_0 \in A_\delta$) generator of $\mathscr H$ passing through $p_0$. Note that $\mathbf n_i$ is $C^1$. Let $\mathscr G_i$ be the union of the geodesics starting from $\widehat N_i$ with initial velocities given by $\mathbf n_i$. 
Let $\gamma \colon [0, t] \to \mathscr G_i$ denote the curve $s \mapsto \beta_s(p_0)$. 
We wish to choose a subset of $\mathscr G_i$ which is a $C^2$ hypersurface containing $\gamma$. Define $\exp \colon \Omega \to M$ by $\exp(\tau, q) = \exp_q(\tau\mathbf n_i(q))$, where $\Omega \subseteq \R \times \widehat N_i$ is the largest subset on which $\exp$ may be defined. Proposition A.3 of \cite{CDGH01} says that if $\exp_*$ is injective at $(\tau, q)$ then there is an open neighborhood $\mathcal O$ of $(\tau, q)$ such that $\exp(\mathcal O)$ is a $C^2$ submanifold of $M$. This, together with the fact that $\exp$ is injective when restricted to $[0, t] \times \{p_0\}$, shows that some neighborhood of $\gamma$ in $\mathscr G_i$ is a $C^2$ hypersurface in $M$.
Hence we need to show that $\exp_*$ is injective at $(s, p_0)$ for each $s \in [0, t]$.
Note that $\widehat N_i$ is a $C^2$ spacelike submanifold of $M$ of codimension $2$, and that $\widehat N_i$ is \emph{second order tangent} to $\mathscr H$ at $p_0$ in the sense of \cite[Section~4.2]{CDGH01} since $D^2g_i(x_0) = D^2f(x_0)$. By \cite[Lemma~4.15]{CDGH01} there can then be no focal point of $\widehat N_i$ along $\gamma$.
By \cite[Proposition 30, Chapter 10]{ONeill} this means that $\exp_*$ is injective at $(s, p_0)$ for all $s \in [0, t]$.
As pointed out previously, \cite[Proposition A.3]{CDGH01} then tells us that some open neighborhood of $\gamma$ in $\mathscr G_i$ is a $C^2$ submanifold. Since $\exp_*$ is injective at each point on $\gamma$, it is injective in a neighborhood of each such point. Since $\gamma([0, t])$ is compact, finitely many such neighborhoods suffice to cover $\gamma$. Hence there is a neighborhood of $\gamma$ where $\mathscr G_i$ is $C^2$ and $\exp_*$ is injective. Denote this neighborhood by $\mathscr H_i$.

By an application of \autoref{Null geodesically spanned hypersurfaces are null} we see that $\mathscr H_i$ is a null hypersurface.
\end{proofpart}

\begin{proofpart}{Definition of a map $\widehat{\beta_t}^i \colon \mathscr H_i \to M$}
By definition of $\mathscr H_i$, the vector field $\mathbf n_i$ (where defined) is tangent to $\mathscr H_i$. Since $\mathscr H_i$ is a null hypersurface, it has a unique $\sigma$-unit normal null vector field, which must then be an extension of $\mathbf n_i$. Call this extension $\mathbf n_i$ as well.

Note that $\mathscr H_i$ contains both $\widehat N_i$ and the generator passing through $p_0$, these being submanifolds of $N_i$ transverse to each other. Hence the first and second derivatives of the graphing functions of $\mathscr H_i$ and $N_i$ must agree at $p_0$. 

Define the map $\widehat{\beta_t}^i \colon \mathscr H_i \to M$ by
\[\widehat{\beta_t}^i(q) = \exp^g(r(q)\mathbf n_i)\]
where $r(q)$ is the unique nonnegative real number such that the $\sigma$-distance from $q$ to $\exp^g(r(q)\mathbf n_i)$ along the $g$-geodesic $\exp^g(\tau \mathbf n_i)$ is precisely $t$. Then by definition
\[\widehat{\beta_t}^i(p_0) = \widehat{\beta_t}(p_0).\]
Note that $\widehat{\beta_t}$ and $\widehat{\beta_t}^i$ are defined by the formula $\exp^g(r(q)\mathbf k)$ where $\mathbf k$ is the normal vector field of $N$ and $\mathscr H_i$, respectively. The derivative of $\exp^g(r(q)\mathbf k)$ is determined by the first derivatives of $r$ and $\mathbf k$. These in turn are determined by the second derivatives of the graphing functions of $N$ and $\mathscr H_i$.
Since $dg(p_0) = dg_i(p_0)$ and $D^2g(p_0) = D^2g_i(p_0)$ this means that the tangent maps of $\widehat{\beta_t}^i$ and $\widehat{\beta_t}$ agree at $p_0$.
Hence
\[J(\widehat{\beta_t}^i)(p_0) = J(\widehat{\beta_t})(p_0).\]
We have now reduced the problem to showing that $J(\widehat{\beta_t}^i)(p_0) \geq 1$.
\end{proofpart}

\begin{proofpart}{Computation of $J(\widehat{\beta_t}^i)(p_0)$}
Let $b_{\mathscr H_i}$ be the one-parameter family of Weingarten maps (defined in \autoref{The null Weingarten map}) along the generator of $\mathscr H_i$ through $p_0$ with its affine parametrization, and let $\dot b_{\mathscr H_i}$ denote the covariant derivative of $b$ along this generator. Equation \eqref{Riccati equation for null Weingarten map} in \autoref{The null Weingarten map} tells us that $b_{\mathscr H_i}$ satisfies the equation
\begin{equation}\label{Riccati equation along a generator}
\dot b + b^2 + \widetilde R = 0.
\end{equation}
Recall from \autoref{Alexandrov Riccati equation} that $\mathscr H$ has a null Weingarten map $b\Al$, defined in terms of Alexandrov derivatives, on all points to the past of $p_0$ on the generator of $\mathscr H$ through $p_0$, and that this map also satisfies equation \eqref{Riccati equation along a generator}.
Since the null Weingarten map of a null hypersurface can be expressed in the first two derivatives of a graphing function, and $\mathscr{H}$ shares these derivatives with $N_i$ which in turn shares them with $\mathscr H_i$, it holds that $b_{\mathscr H_i}(p_0) = b\Al(p_0)$. By the uniqueness of solutions to the ordinary differential equation \eqref{Riccati equation along a generator}, these two maps must agree on all of the past of $p_0$ along the generator through $p_0$. Let $\theta_{\mathscr H_i}$ denote the null mean curvature of $\mathscr H_i$, as defined in \autoref{The null Weingarten map}. Then we have
\[\theta_{\mathscr H_i} = \tr b_{\mathscr H_i} = \tr b\Al = \theta\Al.\]
\autoref{Integral expression for J} implies that
\[J(\widehat{\beta_t}^i)(p_0) = \exp\left(-\int_0^t \theta_{\mathscr H_i}(\widehat{\beta_s}^i(p_0))\,ds\right).\]
Since $J(\widehat{\beta_t}^i)(p_0) = J(\widehat{\beta_t})(p_0)$ and $\theta\Al = \theta_{\mathscr H_i}$ along the curve $s \mapsto \widehat{\beta_s}^i(p_0) = \widehat{\beta_s}(p_0)$, we then know that
\[J(\widehat{\beta_t})(p_0) = \exp\left(-\int_0^t \theta\Al(\widehat{\beta_s}(p_0))\,ds\right).\]
Recall that $g(\widehat B)$ has full $\mathfrak h^n$-measure in $N \cap U$ and that $p_0$ was an arbitrary point of $g(\widehat B)$. Since we have assumed that $\theta\Al \leq 0$, we can conclude that $J(\widehat{\beta_t}) \geq 1$ almost everywhere on the neighborhood $U$ of $p$. This completes the proof of the first part of the lemma.
\end{proofpart}

To prove the last part of the lemma, some measure theoretical technicalities remain.

\begin{claim}{$J(\widehat{\beta_t}) = 1$ almost everywhere on $\tilde A_\delta \cap C$ only if $\theta\Al = 0$ almost everywhere on $\tilde A_\delta \cap C$}
Recall that $C$ is an arbitrary closed subset of $U$.
Suppose that $J(\widehat{\beta_t}) = 1$ almost everywhere on $\tilde A_\delta \cap C$ with respect to $\mathfrak h^n$. Then in particular it holds that $J(\widehat{\beta_t}) = 1$ almost everywhere on $\tilde A_\delta \cap C \cap g(\widehat B)$. (Recall that $g(\widehat B)$ has full measure in $N \cap U$.) We have seen that on this set
\[J(\widehat{\beta_t})(q) = \exp\left(-\int_0^t \theta\Al(\widehat{\beta_\tau}(q))\,d\tau\right).\]
Recall that we have assumed that $\theta\Al \leq 0$.
This means that if $J(\widehat{\beta_t}) = 1$ almost everywhere, then it holds for $\mathfrak h^n$-almost every $q \in \tilde A_\delta \cap C \cap g(\widehat B)$ that
\[\int_0^t \theta\Al(\widehat{\beta_\tau}(q))\,d\tau = 0.\]
In other words
\[\int_{C \cap N} \int_0^t \theta\Al(\widehat{\beta_\tau}(q))\,d\tau \,d\mathfrak h^n(q) = 0.\]
Note that the Hausdorff measure $\mathfrak h^n$ differs only by a $C^1$ function from the Lebesgue measure in coordinates on $C \cap N$. Hence Fubini's theorem \cite[Theorem~2.6.2]{Federer69} applied in coordinates implies that
\[\int_0^t \int_{C \cap N} \theta\Al(\widehat{\beta_\tau}(q)) \,d\mathfrak h^n(q) \,d\tau = 0.\]
In other words, it holds for almost every $\tau \in [0, t]$ that
\[\int_{C \cap N} \theta\Al(\widehat{\beta_\tau}(q)) \,d\mathfrak h^n(q) = 0.\]
This means that
\[\int_{C \cap N} \theta\Al(\widehat{\beta_\tau}(q)) J(\widehat{\beta_\tau})(q) \,d\mathfrak h^n(q) = 0\]
for almost every $\tau \in [0, 1]$, where $J(\widehat{\beta_\tau})$ denotes the determinant of the jacobian of $\widehat{\beta_\tau}$.
Using Theorem~3.1 of \cite{Federer59} we see that
\[\int_{C \cap N} \theta\Al(\widehat{\beta_\tau}(q)) J(\widehat{\beta_\tau})(q) \,d\mathfrak h^n(q) = \int_{\widehat{\beta_\tau}(C \cap N)} \theta\Al(q) \,d\mathfrak h^n(q)\]
so that
\[\int_{C \cap N \cap \widehat{\beta_\tau}(C \cap N)} \theta\Al(q) \,d\mathfrak h^n(q) = 0\]
for almost every $\tau \in [0, t]$. In particular, there is a decreasing sequence $\tau_k \to 0$ such that
\[\int_{C \cap N \cap \widehat{\beta_{\tau_k}}(C \cap N)} \theta\Al(q) \,d\mathfrak h^n(q) = 0\]
Hence $\theta\Al = 0$ almost everywhere on each of the sets $C \cap N \cap \widehat{\beta_{\tau_k}}(C \cap N)$. Since these form a countable increasing sequence with union $C \cap N$, it holds that $\theta\Al = 0$ almost everywhere on $C \cap N$ with respect to $\mathfrak h^n$, which completes the proof.
\end{claim}
\end{proof}
\end{lem}

\begin{prop}\label{Area unchanged by flow}
Let $\mathscr H$ be a horizon in an $(n+1)$-dimensional spacetime $M$ equipped with a Riemannian metric $\sigma$ of the form
\[\sigma(X, Y) = g(X, Y) + 2g(X, V)g(Y, V)\]
for some timelike $g$-unit vector field $V$.
Let $\delta > 0$ and let $A_\delta$ be the $\delta$-flow set of $\mathscr H$. Let $\mathfrak h^n$ denote the $n$-dimensional Hausdorff measure induced by $\sigma$.
Let $\widehat{\mathscr H}$ be a past null geodesically ruled open subset of $\mathscr H$. Suppose that $\theta\Al \leq 0$ on all of $\mathscr H\Al \cap \widehat{\mathscr H}$.
Let $t > 0$ be such that the generator flow $\beta_t$ is defined on all of $\widehat{\mathscr H} \cap A_\delta$. 

Then
\[\mathfrak h^n(\widehat{\mathscr H} \cap A_\delta) = \mathfrak h^n(\beta_t(\widehat{\mathscr H} \cap A_\delta)).\]
Moreover, $\theta\Al = 0$ almost everywhere on $\widehat{\mathscr H} \cap A_\delta$.
\begin{proof}
Note that the generators of $\widehat{\mathscr H}$ are the intersections of the generators of $\mathscr H$ with $\widehat{\mathscr H}$ since $\widehat{\mathscr H}$ is an open subset of the horizon $\mathscr H$.
For each $p$, let $Z_p$ and $\Psi_p$ be the neighborhoods and functions given by \autoref{Local volume change under flow}. Since $A_\delta$ is second countable, a countable number of neighborhoods $Z_1, Z_2, \ldots$ suffice to cover $A_\delta$. 
For each $i \geq 1$, let
\[Y_i = Z_i \setminus \bigcup_{1 \leq j < i} Z_j.\]
Then
\begin{itemize}
\item each $Y_i$ is measurable,
\item $Y_i \subseteq Z_i$ for each $i$,
\item the $Y_i$ are pairwise disjoint,
\item $\bigcup_{i \geq 1} Y_i = \bigcup_{i \geq 1} Z_i \supseteq A_\delta$.
\end{itemize}
For each $i \geq 1$, let $\varphi_i$ be the indicator function of $Y_i \cap \widehat{\mathscr H} \cap \tilde A_\delta$. Then \autoref{Local volume change under flow} says that
\[\int_{\tilde A_\delta} \varphi_i \Psi_i d\mathfrak h^n = \int_{\beta_t(\tilde A_\delta)} \varphi_i(\beta_t^{-1}(y)) d\mathfrak h^n(y)\]
for each $i \geq 1$. Since each $\varphi_i$ is zero outside of $\widehat{\mathscr H}$, this means that
\[\int_{\widehat{\mathscr H} \cap \tilde A_\delta} \varphi_i \Psi_i d\mathfrak h^n = \int_{\beta_t(\widehat{\mathscr H} \cap \tilde A_\delta)} \varphi_i(\beta_t^{-1}(y)) d\mathfrak h^n(y)\]
Taking a sum over $i$, we see that
\[\int_{\widehat{\mathscr H} \cap \tilde A_\delta} \sum_{i \geq 1} \varphi_i \Psi_i d\mathfrak h^n = \int_{\beta_t(\widehat{\mathscr H} \cap \tilde A_\delta)} \sum_{i \geq 0}\varphi_i(\beta_t^{-1}(y)) d\mathfrak h^n(y).\]
Since precisely one of the functions $\varphi_i$ is nonzero at any point $p \in \widehat{\mathscr H} \cap \tilde A_\delta$, and takes the value $1$ there, we have $\sum_{i \geq 1} \varphi_i \Psi_i \geq 1$ almost everywhere on $\widehat{\mathscr H} \cap \tilde A_\delta$. Moreover, we have $\sum_{i \geq 0}\varphi_i(\beta_t^{-1}(y)) = 1$ almost everywhere on $\beta_t(\widehat{\mathscr H} \cap \tilde A_\delta)$. 
Hence
\[\mathfrak h^n(\widehat{\mathscr H} \cap \tilde A_\delta) \leq \int_{\widehat{\mathscr H} \cap \tilde A_\delta} \sum_{i \geq 1} \varphi_i \Psi_i d\mathfrak h^n = \mathfrak h^n(\beta_t(\widehat{\mathscr H} \cap \tilde A_\delta)).\]
Since $\tilde A_\delta$ has full measure in $A_\delta$ and $\beta_t(\tilde A_\delta)$ has full measure in $\beta_t(A_\delta)$ by \autoref{Flowed A_delta has full measure}, this means that
\[\mathfrak h^n(\widehat{\mathscr H} \cap A_\delta) \leq \mathfrak h^n(\beta_t(\widehat{\mathscr H} \cap A_\delta)).\]
However, $\beta_t(\widehat{\mathscr H} \cap A_\delta) \subseteq \widehat{\mathscr H} \cap A_\delta$ since the generators of $\widehat{\mathscr H}$ agree with the generators of $\mathscr H$, so we also know that
\[\mathfrak h^n(\widehat{\mathscr H} \cap A_\delta) \geq \mathfrak h^n(\beta_t(\widehat{\mathscr H} \cap A_\delta))\]
by additivity of the measure. Hence equality must hold, and the proof of the first statement is complete.

Equality can hold only if $\sum_{i \geq 1} \varphi_i \Psi_i = 1$ almost everywhere on $\widehat{\mathscr H} \cap A_\delta$. This means that each function $\Psi_i$ must be equal to $1$ almost everywhere on $Y_i \cap \widehat{\mathscr H} \cap A_\delta$. By \autoref{Local volume change under flow} this implies that $\theta\Al = 0$ almost everywhere on $Y_i \cap \widehat{\mathscr H} \cap A_\delta$. Since these sets cover $\widehat{\mathscr H} \cap A_\delta$, we have shown that $\theta\Al = 0$ almost everywhere on $\widehat{\mathscr H} \cap A_\delta$ with respect to the measure $\mathfrak h^n$. This completes the proof.
\end{proof}
\end{prop}

\subsubsection{Smoothness from area-preserving generator flow}
\begin{prop}\label{No endpoints}
Let $\mathscr H$ be a horizon in a spacetime of dimension $n+1$ equipped with a Riemannian metric $\sigma$ and the corresponding Hausdorff measure $\mathfrak h^n$.
Let $\Omega$ be a past null geodesically ruled open subset of $\mathscr H$.
Let $A_\delta$ denote the $\delta$-flow set of $\mathscr H$ with respect to $\sigma$.
Suppose that
\[\mathfrak h^n(\Omega \cap A_\delta) = \mathfrak h^n(\beta_t(\Omega \cap A_\delta))\]
for all $t > 0$ and all $\delta > 0$, and that $\mathfrak h^n(\Omega) < \infty$. 

Then the following two statements hold.
\begin{enumerate}
\item The union of the images of generators which are inextendible and completely contained in $\Omega$ is a dense subset of $\Omega$.
\item No generator of $\mathscr H$ has any endpoint on $\Omega$.
\end{enumerate}
\begin{proof}
Let $A_\delta$ denote the $\delta$-flow set of $\mathscr H$ with respect to $\sigma$. Recall that the total flow set of $\mathscr H$ is the set
\[A_0 = \bigcup_{\delta > 0} A_\delta.\]
Note that if $\delta < \delta'$ then $A_\delta \supseteq A_{\delta'}$. Hence
\[A_0 = \bigcup_{\delta > 0} A_\delta = \bigcup_{k \in \Z^+} A_{1/k}.\]
Since the family $\Omega \cap A_{1/k}$ is increasing,
\[\mathfrak h^n(\beta_t(\Omega \cap A_0)) = \lim_{k \to \infty} \mathfrak h^n(\beta_t(\Omega \cap A_{1/k})) = \lim_{k \to \infty} \mathfrak h^n(\Omega \cap A_{1/k}) = \mathfrak h^n(\Omega \cap A_0).\]
The limits are finite, since by hypothesis $\mathfrak h^n(\Omega \cap A_{1/k})$ is uniformly bounded with respect to $k$ by $\mathfrak h^n(\Omega)$.

Introduce the sets $\mathfrak C$ and $\mathfrak D$ defined by
\[\mathfrak C = \bigcap_{t \in \Z^+} \beta_t(\Omega \cap A_0),\]
\[\begin{split}
\mathfrak D = \{p \in \Omega \mid &\text{ there is a unique generator through $p$,}\\
				&\text{ and this generator has no future endpoint}\}.
\end{split}\]
We first show that if $p \in \mathfrak C$ then the generator $\Gamma_p$ through $p$ is an inextendible geodesic contained in $\Omega$. By choice of $\Omega$, the part of the generator to the past of $p$ belongs to $\Omega$, and is inextendible in the past direction. Parameterize $\Gamma_p$ by an affine parameter such that $\Gamma_p(0) = p$. Suppose that the maximal future extension of the generator were to leave $\Omega$ at some point $q = \Gamma_p(s)$. Since $\Gamma_p$ is smooth and the interval $[0, s]$ is compact, the curve segment $\Gamma_p([0, s])$ has finite length in the Riemannian metric $\sigma$. This means that $p \notin \beta_t(\Omega \cap A_0)$ whenever $t$ is greater than this length, contradicting the assumption that $p \in \mathfrak C$. This shows that the set $\mathfrak C$ satisfies the conditions for the first statement in the conclusion, so that it is sufficient to show that $\mathfrak C$ is dense to complete the proof of that statement.

Note that the fact that generators through points of $\mathfrak C$ are inextendible means that they can have no endpoints. Hence $\mathfrak C \subseteq \mathfrak D$.
Since $(\beta_t(\Omega \cap A_0))_{t = 1}^\infty$ is a countable decreasing family of sets of equal measure it holds that $\mathfrak h^n(\mathfrak C) = \mathfrak h^n(\Omega \cap A_0)$.
Since $A_0 \subseteq \mathscr H$, and $\mathfrak h^n(\mathscr H \setminus A_0) = 0$ by \autoref{Union of flow sets has full measure}, this means that $\mathfrak h^n(\Omega) = \mathfrak h^n(\Omega \cap A_0)$ so that
\[\mathfrak h^n(\mathfrak C) = \mathfrak h^n(\Omega).\]
In particular, $\mathfrak C$ is dense in $\Omega$. 
Since $\mathfrak C \subseteq \mathfrak D$, it follows that $\mathfrak D$ is also dense in $\Omega$.

We will now show that $\mathfrak D$ is closed in $\Omega$. Suppose that a sequence $(p_k)_{k \in \N}$ in $\mathfrak D$ converges to $p \in \Omega$. Let $X_k$ denote the (unique, by definition of $\mathfrak D$) future-directed $\sigma$-unit tangent of a generator at $p_k$. The $\sigma$-unit tangent bundle over the compact countable set $\{p\} \cup \{p_1, p_2, \ldots\}$ is compact, so by passing to a subsequence we may assume that the $X_k$ converge to some unit vector $X$ at $p$. By \cite[Lemma~6.4]{CDGH01} the space of past-directed $\sigma$-unit generator tangents is closed in the unit tangent bundle, so we know that $X$ is tangent to a generator. Let $\gamma$ denote the inextendible geodesic with initial velocity $X$, and let $\gamma_k$ denote the inextendible geodesic with initial velocity $X_k$. By definition of $\mathfrak D$, each geodesic $\gamma_k$ avoids the open set $I^+(\mathscr H)$. By continuous dependence on initial conditions for ordinary differential equations, $\gamma$ must also avoid the open set $I^+(\mathscr H)$. Suppose to get a contradiction that $\gamma$ leaves $\mathscr H$ at some point $p$. Choose coordinates around $p$ of the form $(-a, a) \times \Sigma$, such that each curve $(-a, a) \times \{q\}$ is timelike and each slice $\{t\} \times \Sigma$ is spacelike. Let $\Sigma'$ denote the projection of $\im \gamma \cap ((-a, a) \times \Sigma)$ on $\Sigma$. Recall that $\mathscr H$ is an achronal hypersurface, so by possibly shrinking $\Sigma$ we may represent $\mathscr H \cap ((-a, a) \times \Sigma')$ as the graph of a function $f_{\mathscr H} \colon \Sigma' \to (-a, a)$. Let $f_\gamma \colon \Sigma' \to (-a, a)$ be the function the graph of which is the image of $\gamma$ (on both sides of the point $p$). Recall that $\gamma$ is a null curve, so if $f_{\mathscr H}(x) > f_\gamma(x)$ at some point $x \in \Sigma'$ then there is a timelike curve from $(f_{\mathscr H}(x), x)$ to $p$, contradicting achronality of $\mathscr H$. If $f_{\mathscr H}(x) < f_\gamma(x)$ at some point $x \in \Sigma'$ then $\gamma$ intersects $I^+(\mathscr H)$ which we saw earlier is impossible. 
Hence $\gamma$ cannot leave $\mathscr H$ to the future, and so there is a generator through $p$ without future endpoint. Moreover, $p$ is then an interior point of a generator so this generator is unique. Hence $p \in \mathfrak D$ and we have shown that $\mathfrak D$ is closed in $\Omega$.

We have now shown that $\mathfrak D$ is a closed dense subset of $\Omega$. Hence $\mathfrak D = \Omega$.
Since no point in $\mathfrak D$ lies on a generator with a future endpoint, no point in $\mathfrak D$ can be a future endpoint of a generator.
This shows that no generator of $\mathscr H$ can have a future endpoint on $\Omega$. Recall that no generator has any past endpoint either, since $\mathscr H$ is a horizon. This completes the proof. 
\end{proof}
\end{prop}

The condition of $\mathscr H$ containing no endpoints is very strong when combined with the condition \(\theta\Al = 0\), as the following theorem due to Chru\'sciel, Delay, Galloway and Howard (see \cite[Theorem~6.18]{CDGH01}) illustrates.
\begin{thm}\label{No endpoints implies smoothness}
Suppose that $\Omega$ is an open subset of a horizon $\mathscr H$ in a spacetime $M$, such that $\Omega$ contains no endpoints of generators of $\mathscr H$. Suppose moreover that $\theta\Al = 0$ almost everywhere with respect to the $n$-dimensional Hausdorff measure $\mathfrak h^n$ induced by a Riemannian metric $\sigma$ on $M$.
Then $\Omega$ is a smooth submanifold of $M$. Moreover, if the metric on $M$ is analytic then $\Omega$ is an analytic submanifold of $M$.
\end{thm}

We are now in a position to prove our main theorem.
It was shown in \cite[Section~4]{BudzynskiKondrackiKrolak03} that not all compact horizons are smooth. Our theorem shows that the additional hypothesis of the null energy condition is sufficient to guarantee smoothness. Note that an analogous result holds for past Cauchy horizons, as can be seen by reversing the time orientation.
\begin{thm}\label{Compact horizons are smooth}
Let $M$ be a spacetime of dimension $n+1$ satisfying the null energy condition. Let $S \subset M$ be an acausal set with $\edge(S) = \emptyset$.
Let $\widehat{\mathscr H}$ be an open subset of $H^+(S)$ with compact closure. Suppose that $\widehat{\mathscr H}$ is past null geodesically ruled.
Then $\widehat{\mathscr H}$ is a smooth, totally geodesic, null hypersurface. If moreover the metric is analytic, then $\widehat{\mathscr H}$ is an analytic hypersurface.
\begin{proof}
First note that \autoref{Cauchy horizons are horizons} tells us that $H^+(S)$ is a horizon in the sense of \autoref{Definition of horizon}. Further note that $\widehat{\mathscr H}$ is a Lipschitz hypersurface, since it is an open subset of the Lipschitz hypersurface $H^+(S)$. It is also assumed to be past null geodesically ruled, so the generators of $\widehat{\mathscr H}$ are the intersection of the generators of $H^+(S)$ with $\widehat{\mathscr H}$. Note that Alexandrov points of $H^+(S)$ are Alexandrov points of $\widehat{\mathscr H}$.
Each generator of $\widehat{\mathscr H}$ is a part of a null geodesic contained in $H^+(S)$. By definition, each generator of $\widehat{\mathscr H}$ is completely contained in, and hence totally past imprisoned in, the compact set $\overline{\widehat{\mathscr H}}$. This means that the generator flow $\beta_t$ is defined for all $t$ on the part of the total flow set of $H^+(S)$ which lies in $\widehat{\mathscr H}$.
Moreover, we can show using \autoref{H+complete} that almost every point of $\widehat{\mathscr H}$ belongs to a generator which is complete in the past direction.
To apply that lemma to a generator, we must show that the intersection of the generator with a sufficiently small spacetime neighborhood of any of its points is contained in a $C^{1,1}$ hypersurface. However, the flow set $A_0$ of the horizon contains full generators, and each point such point belongs to some $A_\delta$ for some sufficiently small $\delta > 0$. (To define $A_\delta$ we use a Riemannian metric, but it does not matter precisely which metric is chosen.) By \autoref{N construction}, the intersection of $A_\delta$ with some spacetime neighborhood of any point of $A_\delta$ belongs to a $C^{1, 1}$ hypersurface. This means that \autoref{H+complete} is applicable to all generators contained in $A_0$, so that all points on $A_0$ belong to a generator which is complete in the past direction. Since the null energy condition holds, this means that \autoref{Sign of theta} is applicable, telling us that $\theta\Al \leq 0$ on $A_0$, which has full measure in $\mathscr H$. Hence $\theta\Al \leq 0$ almost everywhere on $\mathscr H$.
Let $V$ be an arbitrary unit timelike vector field on $M$, introduce the Riemannian metric $\sigma$ on $M$ defined by
\begin{equation}\label{Riemannian metric from vector field}
\sigma(X, Y) = g(X, Y) + 2g(X, V)g(Y,V),
\end{equation}
and let $\mathfrak h^n$ be the corresponding $n$-dimensional Hausdorff measure. 
By \autoref{Subsets of compact sets have finite measure}, the set $\widehat{\mathscr H}$ and all its measurable subsets have finite $\mathfrak h^n$-measure since $\overline{\widehat{\mathscr H}}$ is compact.
For each $\delta > 0$, let $A_\delta$ denote the $\delta$-flow set of $H^+(S)$. By \autoref{Area unchanged by flow} we know that
\[\mathfrak h^n(\widehat{\mathscr H} \cap A_\delta) = \mathfrak h^n(\beta_t(\widehat{\mathscr H} \cap A_\delta))\]
and
\[\theta\Al = 0 \text{ almost everywhere on } \widehat{\mathscr H} \cap A_\delta\]
for all $t > 0$ and all $\delta > 0$.
\autoref{No endpoints} then tells us that no generator of $H^+(S)$ has any endpoint on $\widehat{\mathscr H}$. 
Moreover, since the total flow set $A_0 = \bigcup_{\delta > 0} A_\delta$ has full $\mathfrak h^n$-measure by \autoref{Union of flow sets has full measure} we see that $\theta\Al = 0$ almost everywhere on $\widehat{\mathscr H}$.
\autoref{No endpoints implies smoothness} then says that $\widehat{\mathscr H}$ is a smooth submanifold of $M$, and that it is analytic if the metric is analytic. Since $\widehat{\mathscr H}$ is an open subset of $H^+(S)$ and the tangent space of $H^+(S)$ is a null hyperplane whenever it exists, $\widehat{\mathscr H}$ is a null hypersurface.

Let $K$ be a tangent vector field of the generators of $\widehat{\mathscr H}$.
Since $\widehat{\mathscr H}$ is smooth, its null mean curvature $\theta$ with respect to $K$ is a smooth function and its sign agrees with that of the Alexandrov null mean curvature $\theta\Al$. We saw previously that $\theta\Al = 0$ almost everywhere, so by continuity $\theta = 0$ everywhere. 
Let $b$ denote the null Weingarten map with respect to $K$, and let $S = b - \frac{\theta}{n-2}$. Since $S$ is self-adjoint, $\tr(S^2) \geq 0$. Since $\theta = 0$ and $\Ric(K, K) \geq 0$ everywhere, equation \eqref{Raychaudhuri equation} tells us that $\tr(S^2) = 0$. Since $S$ is self-adjoint this implies that $S = 0$. Hence $b = 0$ everywhere, so that $\widehat{\mathscr H}$ has everywhere zero null second fundamental form. \autoref{Vanishing null second fundamental form implies totally geodesic hypersurface} then implies that $\widehat{\mathscr H}$ is totally geodesic, completing the proof.
\end{proof}
\end{thm}
The following corollary is immediate.
\begin{cor}
Let $M$ be a spacetime satisfying the null energy condition. Let $S \subset M$ be an acausal set with $\edge(S) = \emptyset$. 
Suppose that $H^+(S)$ is compact. Then $H^+(S)$ is a smooth, totally geodesic, null hypersurface. If moreover the metric is analytic then $H^+(S)$ is an analytic hypersurface.
\begin{proof}
Let $\widehat{\mathscr H} = H^+(S)$ and apply \autoref{Compact horizons are smooth}.
\end{proof}
\end{cor}

\begin{remark}\label{Stronger results}
The methods used here are sufficient to prove a slightly stronger statement. The hypothesis that the horizon is a Cauchy horizon, rather than an arbitrary horizon in the sense of \autoref{Definition of horizon}, is used only to prove that its generators are complete in the past direction. If one were to know for some other reason that the generators are complete in the past direction, then the result can be applied to general horizons.\\
Moreover, the hypothesis that the horizon is compact is used only to prove that the generators are complete in the past direction and to ensure that the horizon has finite measure in the $n$-dimensional Hausdorff measure associated to the Riemanniann metric defined in \eqref{Riemannian metric from vector field} using an arbitrary unit timelike vector field $V$. This means that if it is known that the generators are complete in the past direction and that there is a vector field $V$ which gives the horizon finite $n$-dimensional Hausdorff measure, then both the compactness hypothesis and the hypothesis that the horizon is a Cauchy horizon may be dropped.
\end{remark}

\section{Lorentzian cobordisms satisfying energy conditions are trivial}\label{Lorentzian cobordisms satisfying energy conditions are trivial}
\subsection{Lorentzian pseudocobordisms}
In this section, we will define the concept of a Lorentzian pseudocobordism, and the various related notions we will use.

\begin{defn}
Let $S$ and $S'$ be smooth, compact manifolds of dimension $n$. A \emph{cobordism} between $S$ and $S'$ is a compact $(n+1)$-dimensional manifold-with-boundary $M$, the boundary of which is the disjoint union $S \sqcup S'$. 
If there is a cobordism between $S$ and $S'$, we say that they are \emph{cobordant}.
\end{defn}
We note in passing that two compact manifolds without boundary are cobordant if and only if their Stiefel-Whitney numbers agree (see \cite[Corollary~4.11]{MilnorStasheff}). In particular (see \cite[p.\,203]{MilnorStasheff}) any two compact three-dimensional manifolds without boundaries are cobordant.

We will need several different notions of Lorentzian cobordisms. Since the word ''cobordism'' generally refers to a compact space, we will define the notion of a ''Lorentzian pseudocobordism'':
\begin{defn}
Let $S_1$ and $S_2$ be manifolds of dimension $n$ without boundary. A \emph{Lorentzian pseudocobordism} between $S_1$ and $S_2$ is a Lorentzian $(n+1)$-manifold $M$, the boundary of which is the disjoint union $S_1 \sqcup S_2$, such that $S_1$ and $S_2$ are spacelike, and $M$ admits a (nowhere zero) timelike vector field which is inward-directed on $S_1$ and outward-directed on $S_2$.
\end{defn}

The classical notion of a Lorentzian cobordism is the following.
\begin{defn}\label{Definition of Lorentzian cobordism}
A Lorentzian pseudocobordism $M$ between $S_1$ and $S_2$ is a \emph{compact Lorentzian cobordism} (or simply \emph{Lorentzian cobordism}) if $M$ is compact.
\end{defn}

It turns out, as was noted by Borde (see \cite{Borde94}), that many of the theorems about Lorentzian cobordisms continue to hold when the property of compactness is replaced by the property of ''causal compactness''. We will call the resulting object a ''causally compact Lorentzian cobordism''.
\begin{defn}\label{Definition of causal compactness}
A spacetime $M$ (with possibly nonempty boundary) is \emph{causally compact} if $\overline{I(p)}$ is compact for each $p \in M$.
\end{defn}
Causal compactness captures the concept of ''compact in time'', while allowing the spacetime to be non-compact in the spatial directions.
\begin{defn}\label{Definition of causally compact Lorentzian pseudocobordism}
A Lorentzian pseudocobordism $M$ between $S_1$ and $S_2$ is called a \emph{causally compact Lorentzian pseudocobordism} if $M$ is causally compact.
\end{defn}
Of course, we immediately see that every (compact) Lorentzian cobordism is also a causally compact Lorentzian pseudocobordism.

\begin{defn}
A Lorentzian pseudocobordism $M$ between $S_1$ and $S_2$ is \emph{topologically trivial} if it is diffeomorphic to $S_1 \times [0, 1]$.
\end{defn}

\subsection{Tipler's theorem}
A theorem due to Geroch \cite[Theorem~2]{Geroch67} states that a topologically nontrivial Lorentzian cobordism cannot satisfy the chronology condition. A result from 1977 by Tipler \cite[Theorems 3 and 4]{Tipler77} further implies that a nontrivial Lorentzian cobordism cannot satisfy certain energy conditions. Unfortunately Tipler's original proof, the methods of which are also used in \cite[p.\,295-298]{HawkingEllis} for proving Hawking's singularity theorem, is flawed in that it is implicitly assumed that a certain Cauchy horizon is $C^2$. In this section, we will apply \autoref{Compact horizons are smooth} to prove Tipler's theorem (here stated as \autoref{Full Tipler theorem}) without needing this assumption.

The following is the theorem about cobordisms which is proved (but not stated in this form) by Tipler in \cite[Theorems 3 and 4]{Tipler77}. Tipler did not mention the need for the condition that $H^+(S_1)$ is $C^2$. However, his proof works when this hypothesis is added.
\begin{thm}\label{Original Tipler theorem}
Let $n \geq 2$, let $S_1$, $S_2$ be compact $n$-dimensional manifolds and let $(M, g)$ be a compact connected Lorentzian cobordism between $S_1$ and $S_2$ which satisfies the either the strict null energy condition or the null energy condition together with the lightlike generic condition. \textbf{Suppose moreover that $H^+(S_1)$ is $C^2$.} Then there exists a diffeomorphism $\varphi \colon S_1 \times [0, 1] \to M$ such that the submanifold $\varphi(\{x\} \times [0, 1])$ is $g$-timelike for every $x \in S_1$; in particular, $S_1$ is diffeomorphic to $S_2$.
\end{thm}

\subsection{Tipler's theorem without smoothness hypothesis}
We will prove the generalization of Tipler's theorem, suggested in \cite{Borde94}, to causally compact Lorentzian pseudocobordisms. 
\begin{thm}\label{Full Tipler theorem}
Let $n \geq 2$. Let $S_1$ and $S_2$ be $n$-dimensional manifolds and let $(M, g)$ be a connected, causally compact Lorentzian pseudocobordism between $S_1$ and $S_2$ which satisfies either the strict null energy condition (i.e. that $\Ric(X, X) > 0$ for all lightlike vectors $X$), or the null energy condition (i.e. that $\Ric(X, X) \geq 0$ for all lightlike vectors $X$) together with the lightlike generic condition (i.e. that each lightlike geodesic $\gamma$ contains at least one point at which $\dot\gamma^e \dot\gamma^f \dot\gamma_{[a} R_{b]ef[c} \dot\gamma_{d]} \neq 0$).

Then $M$ is globally hyperbolic. In particular $M \cong S_1 \times [0, 1]$ so that $S_1$ and $S_2$ are diffeomorphic.
\begin{proof}
Extend $M$ to a manifold without boundary $\hat{M}$ by glueing copies of $S_1 \times [0, \epsilon)$ and $S_2 \times [0, \epsilon)$ to the respective boundaries. 
Consider the future Cauchy horizon $H^+(S_1)$ of $S_1$ in $\hat{M}$. If it is empty, then $S_1$ would be a Cauchy surface for $M$ which would mean that $M$ is globally hyperbolic. Hence it is sufficient to show that $H^+(S_1)$ is empty. Suppose for contradiction that this is not the case.

\begin{claim}{$H^+(S_1)$ is a horizon}
The set $S_1$ is a smooth hypersurface in $\hat{M}$, closed as a set. Note that it is also acausal since a causal curve intersecting $S_1$ more than once would have to do so with the wrong time-orientation. Hence we can use \autoref{Closed spacelike hypersurfaces are edgeless} and \autoref{Cauchy horizons are horizons} to conclude that $H^+(S_1)$ is a horizon in the sense of \autoref{Definition of horizon}.
\end{claim}

Choose a point $p \in I^+(H^+(S_1), M)$. To see that such a point exists, note that $I^+(q, M)$ is nonempty if $q \notin S_2$, and that $H^+(S_1) \setminus S_2$ is nonempty since $H^+(S_1)$ is past null geodesically ruled and hence cannot be completely contained in a spacelike hypersurface. With this choice, $H^+(S_1) \cap I^-(p, M)$ is nonempty.

\begin{claim}{Every generator of $H^+(S_1)$ which intersects $I^-(p, M)$ stays in $I^-(p, M)$ when followed to the past}
Let $\gamma$ be a generator of $H^+(S_1)$. 
We will show that if $\gamma(t) \in I^-(p, M)$ for some $t$ then $\gamma((-\infty, t]) \subseteq I^-(p, M)$. This will mean that $\gamma$ is totally past imprisoned in the set $\overline{I^-(p, M)}$ which is compact since $M$ is causally compact.
To this end, let $\gamma \colon (a, t] \to \hat{M}$ be the maximal past geodesic extension of a generator of $H^+(S_1)$, with $\gamma(t) \in I^-(p)$. 
By \cite[Proposition 53, Chapter 14]{ONeill}, $H^+(S_1)$ and $S_1$ are disjoint, so $\gamma$ does not intersect $S_1$. Moreover, when followed to the past $\gamma$ cannot intersect $S_2$ since it would need to do so with the wrong time orientation.
Hence $\gamma$ stays in $M$. Let $s < t$. Then there is a causal curve in $M$ from $\gamma(s)$ to $p$ formed by concatenating $\gamma$ with a timelike curve from $\gamma(t)$ to $p$. Such a curve exists since we assumed that $\gamma(t) \in I^-(p, M)$. Since this curve is not everywhere lightlike, there is a timelike curve from $\gamma(s)$ to $p$. Hence $\gamma(s) \in I^-(p)$, proving the claim.
\end{claim}

\begin{claim}{$\mathscr H := H^+(S_1) \cap I^-(p, M)$ is past null geodesically ruled}
Let $\mathscr H$ denote the set $H^+(S_1) \cap I^-(p, M)$. 
To find a past complete null geodesic segment through a point $q \in \mathscr H$, consider the intersection of $\mathscr H$ and the generator of $H^+(S_1)$ through $q$. This curve is a geodesic segment, and it is connected and past complete by the previous claim. Hence $\mathscr H$ is past null geodesically ruled.
\end{claim}

\begin{claim}{The existence of $\mathscr H$ is contradictory}
Since $I^-(p, M)$ is open, the set $\mathscr H$ is an open subset of $H^+(S_1)$. Since $M$ is causally compact, the set $\overline{I^-(p, M)}$ is compact. Hence $\mathscr H$ is contained in a compact set and so has compact closure. 
\autoref{Compact horizons are smooth} tells us that $\mathscr H$ is a totally geodesic smooth null hypersurface. The inequality \eqref{Raychaudhuri inequality} from \autoref{The null Weingarten map} then reads
\[\Ric(K, K) \leq 0\]
for all null tangent vectors $K$ to $\mathscr H$. Combining this with the null energy condition, we can conclude that
\[\Ric(K, K) = 0.\]
We will now derive a contradiction from this.

If the spacetime were to satisfy the strict null energy condition, then the contradiction is immediate. 
When we assume that the lightlike generic condition holds, a further argument is needed. 
By \autoref{No endpoints} there is a dense, and in particular nonempty, subset of $\mathscr H$ consisting of points on inextendible null geodesics which are contained in $\mathscr H$. Choose one such inextendible geodesic $\gamma$.
Let $b$ denote the null Weingarten map with respect to a null vector field $K$ which agrees with the tangent vector field of $\gamma$ with an affine parametrization. Since $\mathscr H$ is totally geodesic, $b = 0$ along $\gamma$.
Equation \eqref{Riccati equation for null Weingarten map} then implies that $R(X, K)K$ is parallel to the null vector $K$ for every vector $X$, where $R$ denotes the curvature tensor of the spacetime. By \cite[Proposition 2.2]{BeemHarris93a} this is equivalent to $K^e K^f K_{[a} R_{b]ef[c}K_{d]} = 0$.
However, the lightlike generic condition says that this tensor is nonzero at some point along each inextendible lightlike geodesic. 
We have now obtained a contradiction, so the assumption that the Cauchy horizon $H^+(S_1)$ is nonempty must be false. Hence $M$ is globally hyperbolic.
\end{claim}
\end{proof}
\end{thm}

The above theorem is in one sense the strongest result of this kind one may hope for: The null energy condition is the weakest of the commonly used energy conditions, and in the setting of cobordisms it implies global hyperbolicity, which is the strongest of the commonly used causality conditions.

\subsection*{Acknowledgements}
I am grateful to Mattias Dahl, Piotr Chru\'sciel, Greg Galloway and Marc Nardmann for advice concerning this work.

\appendix
\section{Geometric measure theory}\label{Geometric measure theory}
General references for geometric measure theory are \cite{Federer69} and \cite{Morgan}.

\subsection{Regularity of functions}
\begin{defn}
A function $f \colon \R^n \to \R^m$ is \emph{$C^{1, 1}$} if it is $C^1$ and its differential $df$ is Lipschitz. 
A submanifold of a smooth manifold is $C^{1, 1}$ if it is locally the graph of a $C^{1, 1}$ function in coordinates.
\end{defn}
\begin{defn}\label{Semi-convexity}
A function $f \colon \R^n \to \R$ is \emph{semi-convex} if it is the sum of a convex function and a $C^2$ function.
A submanifold of a smooth manifold is semi-convex if it is locally the graph of a semi-convex function in coordinates.
\end{defn}

\subsection{Measure zero}\label{Measure zero}
Let $\Sigma$ be some smooth manifold of dimension $m$, and let $M$ be a smooth manifold of dimension at least $m$. Let $\psi \colon \Sigma \to M$ be a topological embedding. We will consider two notions of ''measure zero'':
\begin{itemize}
\item Since $\Sigma$ is a smooth manifold (and in particular second countable), any two Riemannian metrics on $\Sigma$ give rise to the same family of sets having measure zero in the associated volume measure on $\Sigma$.
\item Let $\sigma$ be an arbitrary Riemannian metric on $M$. This metric induces a distance function, which in turn induces Hausdorff measures of any dimension. Let $\mathfrak h^m$ denote the $m$-dimensional Hausdorff measure induced by $\sigma$. Then we say that $A \subseteq \psi(\Sigma)$ has measure zero if $\mathfrak h^m(A) = 0$.
\end{itemize}
These two notions are related in the following way.
\begin{prop}\label{Zero sets under Lipschitz maps}
Let $\Sigma$ be some smooth manifold of dimension $m$, and let $M$ be a smooth manifold of dimension $n$ with $n \geq m$. Let $\psi \colon \Sigma \to M$ be a topological embedding. Suppose that $\psi$ is locally Lipschitz. Then $\mathfrak h^m(\psi(A)) = 0$ if $A$ has measure zero viewed as a subset of $\Sigma$. 
\begin{proof}
After representing $\psi$ is coordinates by $\psi \colon U \to V$ with open sets $U \subseteq \Sigma$ and $V \subseteq M$ identified with subsets of $\R^m$ and $\R^n$ it holds that
\[\mathfrak h^m(\psi(A \cap U)) \leq L \mu(A \cap U)\]
where $L$ is the Lipschitz constant of $\psi$ over $U$, and $\mu$ denotes the $m$-dimensional Hausdorff measure on $U$. (This can be proved by bounding the volume change of images of unit balls using the Lipschitz constant, or it can be seen as a special case of the much more powerful \cite[Theorem~2.10.25]{Federer69}.) Since $U$ is a subset of $\R^m$, this Hausdorff measure agrees up to pointwise scaling by a smooth function with the Lebesgue measure in coordinates. In particular, if $A \cap U$ has measure zero in $\Sigma$, then $\mathfrak h^m(\psi(A \cap U)) = 0$. By second countability, countably many charts suffice to cover $\psi(\Sigma)$, and so $\mathfrak h^m(\psi(A \cap U)) = 0$ if $A$ has measure zero.
\end{proof}
\end{prop}

In general, we will mostly be interested in the notions of ''measure zero'' and ''finite measure'', so it will not matter precisely which Riemannian metric is used to induce a measure.

\begin{prop}\label{Subsets of compact sets have finite measure}
Let $M$ be a smooth manifold of dimension at least $n$ and let $N$ be a Lipschitz submanifold of $M$ with dimension $n$. Let $K$ be a compact subset of $N$. Let $\sigma$ be a Riemannian metric on $M$, and let $\mathfrak h^n$ be the corresponding $n$-dimensional Hausdorff measure. Then all measurable subsets of $K$ have finite $\mathfrak h^n$-measure. 
\begin{proof}
It is sufficient to show that $K$ has finite measure, since subsets of $K$ have smaller measure than $K$. Take a finite subcover of the cover $K \subseteq \bigcup_{p \in N} B_\sigma(p, 1)$. Each $N \cap B_\sigma(p, 1)$ has finite $\mathfrak h^n$-measure since $N$ is a Lipschitz hypersurface. Hence we can conclude that $K$ has finite measure.
\end{proof}
\end{prop}

\subsection{Density functions}
A reference for density functions is \cite[Chapter 2]{Morgan}. We will use the same idea, but with somewhat different notation.
\begin{defn}
Let $\mathscr L^n$ denote Lebesgue measure on $\R^n$. For each measurable subset $A \subseteq \R^n$ define the \emph{density function of $A$ (with respect to $\mathscr L^n$)} to be the function
\[\Theta(A, \cdot) \colon A \to [0, 1],\]
\[\Theta(A, q) = \lim_{r \to 0} \frac{\mathscr L^n(A \cap B^n(q, r))}{\mathscr L^n(B^n(q, r)}.\]
Here $B^n(q, r)$ denotes the ball of radius $r$ centered at $q$.
\end{defn}
\begin{defn}\label{Rn full-density subset}
Let $A$ be a measurable subset of $\R^n$. We will call the set
\[\tilde A = \{a \in A \mid \Theta(A, a) = 1\}\]
the \emph{full-density subset} of $A$.
\end{defn}
\begin{prop}\label{Full-density sets in Rn have full measure}
Let $\tilde A$ be the full-density subset of some set $A \subseteq \R^n$. Then $\tilde A$ has full Lebesgue measure in $A$.
\begin{proof}
By \cite[Corollary~2.9]{Morgan} the density function $\Theta(A, \cdot)$ is equal to the characteristic function of $A$ almost everywhere, yielding the conclusion.
\end{proof}
\end{prop}
We now generalize the notion of full-density subsets to hypersurfaces in Riemannian manifolds.
\begin{lem}
Let $(M, \sigma)$ be a Riemannian manifold of dimension $n + 1$ and let $N$ be a Lipschitz hypersurface. Let $U$ be an open subset of $N$ and let $\varphi \colon U \to \R^n$ and $\psi \colon U \to \R^n$ be charts. Let $A$ be a subset of $U$ and let $\tilde A_\varphi$ and $\tilde A_\psi$ denote the full-density subsets of $\varphi(A)$ and $\psi(A)$, respectively. Then
\[\tilde A_\psi = \psi(\varphi^{-1}(\tilde A_\varphi)).\]
\begin{proof}
Abbreviate $\psi \circ \varphi^{-1}$ by $f$. Since $f$ is bi-Lipschitz, it holds for any measurable subset $X$ of $\im \varphi$ that
\[\frac{1}{L_{f^{-1}}^n} \mathscr L^n(X) \leq \mathscr L^n(f(X)) \leq L_f^n \mathscr L^n(X)\]
where $L_{f^-1}$ and $L_f$ denote the Lipschitz constants of $f^{-1}$ and $f$. In particular,
\[\frac{\mathscr L^n(B^n(q, r) \setminus \varphi(A))}{\mathscr L^n(B^n(q, r))} \leq \frac{L_{f^{-1}}^n}{L_f^n} \frac{\mathscr L^n(f(B^n(q, r)) \setminus \psi(A))}{\mathscr L^n(f(B^n(q, r)))}.\]
By letting $R(r)$ be a positive real number such that
\[f(B^n(q, r)) \subseteq B^n(f(q), R(r))\]
and $\rho(r)$ a positive real number such that
\[f(B^n(q, r)) \supseteq B^n(f(q), \rho(r))\]
we see that
\[\frac{\mathscr L^n(B^n(q, r) \setminus \varphi(A))}{\mathscr L^n(B^n(q, r))} \leq \frac{L_{f^{-1}}^n}{L_f^n} \frac{\mathscr L^n(B^n(f(q), R(r)) \setminus \psi(A))}{\mathscr L^n(B^n(f(q), \rho(r)))}.\]
Since $f$ and $f^{-1}$ are Lipschitz, we may choose $R$ and $\rho$ to be bounded from above and below by linear functions of positive derivative, so there are positive constants $D$ and $D'$ such that
\[D \mathscr L^n(B^n(f(q), \rho(r))) \leq \mathscr L^n(B^n(f(q), R(r))) \leq D' \mathscr L^n(B^n(f(q), \rho(r))).\]
This together with the previous inequality means that there is a positive real number $C$ independent of $r$ such that
\[\frac{\mathscr L^n(B^n(q, r) \setminus \varphi(A))}{\mathscr L^n(B^n(q, r))} \leq C \frac{\mathscr L^n(B^n(f(q), R(r)) \setminus \psi(A))}{\mathscr L^n(B^n(f(q), R(r)))}.\]
In particular, if $\Theta(\psi(A), f(q)) = 1$ so that
\[\lim_{r \to 0} \frac{\mathscr L^n(B^n(f(q), R(r)) \setminus \psi(A))}{\mathscr L^n(B^n(f(q), R(r)))} = 0\]
then
\[\lim_{r \to 0} \frac{\mathscr L^n(B^n(q, r) \setminus \varphi(A))}{\mathscr L^n(B^n(q, r))} = 0\]
so that $\Theta(\phi(A), q) = 1$. Hence
\[\Theta(\psi(A), f(q)) = 1\ \implies\ \Theta(\phi(A), q) = 1.\]
Repeating this argument for the inverse of $f$ we see that
\[\Theta(\phi(A), q) = 1\ \implies\ \Theta(\psi(A), f(q)) = 1.\]
This proves that $\tilde A_\psi = f(\tilde A_\varphi)$, completing the proof.
\end{proof}
\end{lem}
\begin{defn}\label{Riemannian full-density subset}
Consider a Riemannian manifold $(M, \sigma)$ of dimension $n + 1$ and let $N$ be a Lipschitz hypersurface. In light of the previous proposition, we may define the \emph{full-density subset} of a set $A \subseteq N$ to be the set $\tilde A$ such that if $\varphi \colon U \to \R^n$ is a chart on $N$ then $\varphi(\tilde A \cap U)$ is the full-density subset of $\varphi(A \cap U)$.
\end{defn}
\begin{defn}
If $q$ belongs to the full-density subset of a set $A$, we say that $q$ is a \emph{full-density point} of $A$.
\end{defn}
\begin{prop}\label{Full-density sets have full measure}
Let $(M, \sigma)$ be a Riemannian manifold of dimension $n + 1$ and let $N$ be a Lipschitz hypersurface. Let $\mathfrak h^n$ be the $n$-dimensional Hausdorff measure constructed from $\sigma$. Let $A \subseteq N$ be any subset and let $\tilde A$ be its full-density subset. Then $\mathfrak h^n(A \setminus \tilde A) = 0$.
\begin{proof}
Since $N$ is second-countable, it is sufficient to prove this locally. This can be done by the use of charts and \autoref{Full-density sets in Rn have full measure}.
\end{proof}
\end{prop}

\begin{prop}\label{Density differentials}
Let $\Omega$ be an open subset of $\R^n$ and let $f, g \colon \Omega \to \R$ be Lipschitz functions. Let $A \subset \Omega$ be a measurable subset of $\Omega$ and suppose that $f$ and $g$ agree on $A$. Let $q$ be a full-density point of $A$ and suppose that $f$ and $g$ are both differentiable at $q$. Then $df(q) = dg(q)$.

If, moreover, $q$ is a point where $f$ and $g$ have second order expansions of the form
\[f(q + \xi) = f(q) + df(q)(\xi) + \frac{1}{2}D^2f(q)(\xi, \xi) + o(|\xi|^2),\]
\[g(q + \xi) = g(q) + dg(q)(\xi) + \frac{1}{2}D^2g(q)(\xi, \xi) + o(|\xi|^2),\]
then $D^2f(q) = D^2g(q)$.
\begin{proof}
Let $h = f - g$ and note that $h$ is differentiable at $q$ and zero on $A$. Suppose that $dh(q)(V) \neq 0$ for some vector $V$ at $q$. By continuity $dh(q)(W)$ for all $W$ in some open neighborhood $U$ of $V$ in $T_q\R^n$. Then for all sufficiently small $\epsilon > 0$ and all $W \in U$ it holds that $h(q + \epsilon U) \neq 0$. This means that $h$ is nonzero on some small open cone in the direction of $V$, which in turn means that $\Theta(A, q)$ cannot be equal to $1$, contradicting the fact that $q$ is a full-density point of $A$. This shows that $dh(q) = 0$, proving the first part of the proposition.

Suppose now that $q$ is a point where $f$ and $g$ have second order expansions in the sense that
\[f(q + \xi) = f(q) + df(q)(\xi) + \frac{1}{2}D^2f(q)(\xi, \xi) + o(|\xi|^2),\]
\[g(q + \xi) = g(q) + dg(q)(\xi) + \frac{1}{2}D^2g(q)(\xi, \xi) + o(|\xi|^2).\]
Then, since $f(q) = g(q)$ and $df(q) = dg(q)$,
\[f(q + \xi) - g(q + \xi) = \frac{1}{2}\left(D^2f(q) - D^2g(q)\right)(\xi, \xi) + o(|\xi|^2).\]
If $\left(D^2f(q) - D^2g(q)\right)(\xi, \xi) = 0$ for every vector $\xi$ then $D^2f(q) - D^2g(q) = 0$ and we are done (since $D^2f(q) - D^2g(q)$ is symmetric), so suppose that there is some $\xi$ with $\left(D^2f(q) - D^2g(q)\right)(\xi, \xi) \neq 0$. By continuity, this then holds for all $\nu$ in a neighborhood of $\xi$, and this means that $f - g$ is nonzero on a small open cone from $q$. This means that $\Theta(A, q)$ cannot be equal to $1$ contradicting the fact that $q$ is a full-density point of $A$, showing that indeed $D^2f(q) = D^2g(q)$.
\end{proof}
\end{prop}

\section{Erratum}
The version of this paper published as \cite{Larsson15} contained some minor errors which have been corrected here.
This appendix summarizes the corrections necessary in \cite{Larsson15}.

\vspace{.5em}
\noindent
\textbf{In the statement of Lemma 1.6:}\\*[.2em]
\begin{tabular}{ll}
Original:  & Let $S$ be an achronal hypersurface\\
Corrected: & Let $S$ be an acausal set with $\edge(S) = \emptyset$
\end{tabular}

\vspace{.5em}
\noindent
\textbf{In the statement of Theorem 1.42:}\\*[.2em]
\begin{tabular}{ll}
Original:  & Let $S \subset M$ be an achronal set with $\edge(S) = \emptyset$.\\
Corrected: & Let $S \subset M$ be an acausal set with $\edge(S) = \emptyset$.
\end{tabular}

\vspace{.5em}
\noindent
\textbf{In the statement of Corollary 1.43:}\\*[.2em]
\begin{tabular}{ll}
Original:  & Let $S \subset M$ be an achronal set with $\edge(S) = \emptyset$.\\
Corrected: & Let $S \subset M$ be an acausal set with $\edge(S) = \emptyset$.
\end{tabular}

\noindent
In addition, the proof of Step X of the proof of Lemma 1.6 in \cite{Larsson15} is incorrect, and should be replaced as indicated in this version.
The proof of Lemma 1.6 in \cite{Larsson15} also contains some typographical errors. These have been corrected here, and some details have been clarified.

\bibliographystyle{abbrv}
\bibliography{References}

\end{document}